%% file: MiniJets.tex
\pdfoutput=1
\documentclass[12pt]{article}
\usepackage{xspace}
\usepackage{cite}

\usepackage{gensymb}
\usepackage{booktabs}
\usepackage{placeins}
\usepackage[unicode=true,pdfusetitle,
 bookmarks=true,bookmarksnumbered=false,bookmarksopen=false,
 breaklinks=false,pdfborder={0 0 1},backref=false,colorlinks=false]
 {hyperref}

\usepackage[pdftex]{graphicx}
\usepackage{amsmath}

\usepackage{color}

\newcommand{\bea}{\begin{eqnarray}}
\newcommand{\eea}{\end{eqnarray}}

\usepackage{color}
\usepackage[makeroom]{cancel}
\usepackage[normalem]{ulem}
\definecolor{pkcolor}{rgb}{0,0.1,0.7}
\definecolor{ascolor}{rgb}{0.7,0.1,0.0}

\newcommand\pkout{\marginpar{\color{pkcolor}$\clubsuit$}\bgroup\markoverwith{\color{pkcolor}{\rule[0.4ex]{2pt}{0.8pt}}}\ULon}

\definecolor{macolor}{rgb}{1,0,1}

\newcommand\maout{\marginpar{\color{macolor}$\heartsuit$}\bgroup\markoverwith{\color{macolor}{\rule[0.4ex]{2pt}{0.8pt}}}\ULon}

\newcommand{\herwig}{\mbox{\textsc{Herwig}}\xspace}
\newcommand{\pythia}{\textsc{Pythia}\xspace}
\newcommand{\sherpa}{\textsc{Sherpa}\xspace}

\newcommand{\ptmin}{\ensuremath{p_{\perp}^{\rm min}}\xspace}
\newcommand{\pdisrupt}{\ensuremath{p_{\rm disrupt}}\xspace}

\newcommand{\ptminnought}{\ensuremath{p_{\perp,0}^{\text{min}}}\xspace}

\newcommand{\preco}{\ensuremath{p_{\rm reco}}\xspace}

\newcommand{\seff}{\ensuremath{\sigma_{\mathrm{eff}}}\xspace}

\begin{document}

\title{Studying minijets and MPI with rapidity correlations \date{}}

\author{M. Azarkin$^1$, P. Kotko$^2$, A. Siodmok$^2$, M. Strikman$^3$
\\\\
{\it \small $^1$ P.N. Lebedev Physics Institute, }\\ {\small\it Moscow 119991, Russia}
\\\\
{\it \small $^2$ Institute of Nuclear Physics PAN}\\ {\small\it Radzikowskiego 152, 31-342 Krak\'ow, Poland}
\\\\
{\it \small $^3$  Department of Physics,   The Pennsylvania State University} \\{\it \small   
University Park, PA 16802, United States}
}
\maketitle



\begin{abstract}
We propose and carry a detailed study of an observable sensitive to different mechanisms of minijet production.
The observables measure how the transverse momenta of hadrons produced in association 
with various trigger objects are balanced as a function of rapidity. 
It is shown that the observables
are sensitive to the model parameters relevant for the minijet production mechanisms: 
low-$p_{\rm T}$ cutoff regulating jet cross-section, transverse distribution 
of partons in protons and parton distribution
functions. We perform our test at different charge-particle
multiplicities and collision energies. The Monte Carlo models, which describe many features of the LHC data, are found
to predict quite different results demonstrating high discriminating power of the proposed observables.
We also review mechanisms and components of \herwig, \pythia, and \sherpa  Monte Carlo models  relevant to 
the minijet production.
\end{abstract} 


\section{Introduction}
\label{Sec:intro}
\input{Introduction.tex}


\section{Observables}
\label{Sec:Observables}
\input{Observables.tex}


\section{Choice of kinematic cuts}
\label{Sec:optimization}
\input{KinematicCuts.tex}


\section{Monte Carlo models}
\label{Sec:MCmodels}
The general purpose Monte Carlo event generators used in our study  
have been reviewed several times, see for example~\cite{Buckley:2011ms,Siodmok:2013zta}. 
Our intention here is not to review them again, but just to provide enough background to set our discussion of
the modelling of minijets.

Before we discuss the event generators, let us however start by recalling briefly of the 
perturbative QCD mechanism of particle production based on the collinear factorization.
In fact, it constitutes the skeleton for all MC event generators. We shall also discuss the modification 
one has to make in the collinear formula to be able to incorporate it into event generators.

\subsection{Minijets in perturbative QCD}
\label{sec:BasicMinijetModel}
\input{BasicMinijetMode.tex}

\subsection{\pythia model}
\label{sec:PythiaModel}
\input{PythiaModel.tex}

\subsection{Herwig model}
\input{HerwigMPImodel.tex}

\subsection{Sherpa model}
\input{Sherpa.tex}


\section{Results}
\label{Sec:Results}
\input{Results.tex}


\section{Summary and conclusions}
\label{Sec:Conclusions}
\input{Summary.tex}

\newpage
\appendix
\input{Appendix.tex}


\FloatBarrier
\bibliography{MiniJets}{}
\bibliographystyle{lucas_unsrt_epjc}
\end{document}

%% file: Introduction.tex
Currently there are number of Monte Carlo (MC) generators which successfully describe many features 
of the inelastic $pp$ collisions at the LHC~\cite{Bahr:2008pv,Bellm:2015jjp,Sjostrand:2014zea,Sjostrand:2006za, Gleisberg:2008ta}. 
Since all MC models assume some physics approximations, it is inevitable that they have 
a number of free parameters which must be fixed by experimental data during the procedure called tuning~\cite{Buckley:2009bj,Khachatryan:2015pea,Skands:2014pea,CMS:2018zub,Seymour:2013qka}. 
It often happens that the description of experimental data by different MC models is similar, despite the fact that
the underlying dynamics in the models differs significantly - with hard collisions giving a major contribution in some models and significant soft contribution in the other.
The aim of this paper is to look for the observables which would be especially sensitive to some 
of the important ingredients  of the models.
Specifically we will propose observables which are sensitive to two important characteristics of the models:
taming of minijet production at small $p_{\rm T}$ and the transverse distribution of partons in the colliding protons.

\begin{figure}
\begin{center}
\includegraphics[ width=0.5\textwidth]{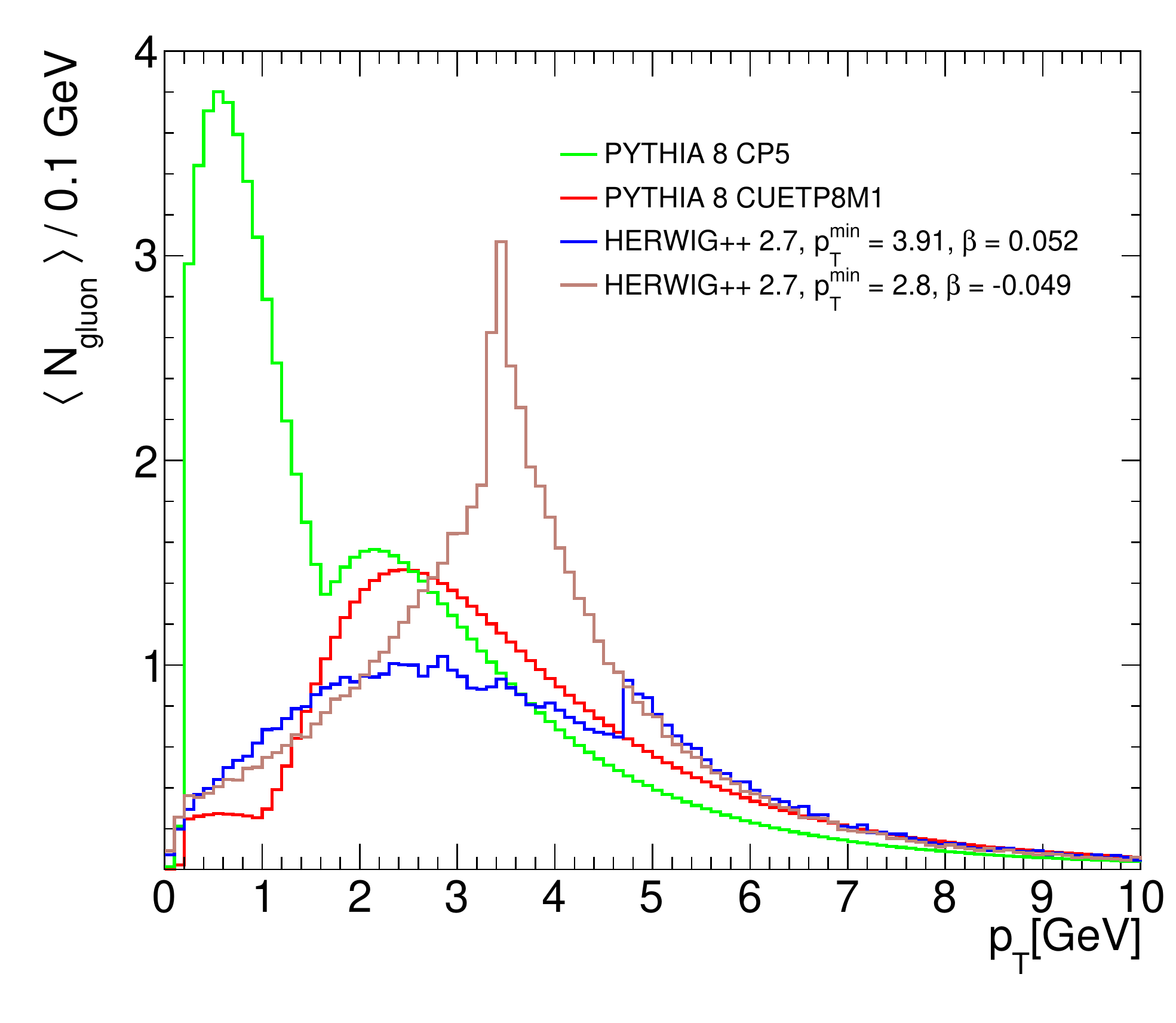}%
\caption{ $p_{\rm T}$ distribution of gluons coming from primary and MPI interactions. 
The pseudorapidity range is $|\eta|<$~5. 
We show two different settings (tunes) for the {\sc Pythia 8}~\cite{Khachatryan:2015pea, CMS:2018zub}
and   {\sc Herwig++ 2.7}~\cite{Seymour:2013qka} MC generators.}%
\label{GluonPtDistribution}%
\end{center}
\end{figure}

Obviously, the rate of parton-parton scattering  has to be tamed at small momentum transfer to avoid 
an unphysical  singular behaviour.
The divergence is usually regulated by including a suppression factor, that is quite different in different models.
Also, in most of the models the suppression for fixed $p_{\rm T}$ becomes stronger with increase of collision energy.
The relevant details of models
used in this study are described in Section~\ref{Sec:MCmodels}.
The number of parton interactions depends not only on the suppression factor, but also on the set of parton distribution 
functions (PDF) and the overlap of the matter distribution of colliding protons. 
It often happens that for some observables models with very different PDF and model parameters are quite close to the data (and to each other). For instance, Fig.~\ref{GluonPtDistribution} shows $p_{\rm T}$
distributions of gluons coming from primary and MPI interactions for successful tunes of {\sc pythia} 
and {\sc herwig}.  One can see that they are very different in the low $p_{\rm T}$ region, 
which eventually produces most of the final-state particles in the collision. 
Nevertheless, the models describe underlying event (UE)~\cite{UE_CMS_7000,UE_ALICE_7000, UE_ATLAS_7000}
and Minimum Bias~\cite{CMS:2018zub} observables 
satisfactory, as the difference in other mechanisms compensates this discrepancy. 
This motivates us to propose  observables which are sensitive to the underlying dynamics of minijet production and, 
thus, allows to discriminate 
models and learn more about underlying dynamics of $pp$ interactions.  The correlation between mechanisms and  their impact on minijet production will be
discussed in Section~ \ref{Sec:MCmodels}.

As mentioned above, the transverse distribution of partons in nucleons is another fitted parameter which is relevant for minijet production. 
Basically, the rate of the double parton interactions (DPS) is inversely proportional to the transverse area occupied by partons. 
This parameter of the models can be conveniently coded 
via so-called sigma effective, $\sigma_{\mathrm{eff}}$,  defined through: $\sigma_{\mathrm{ij}}=\sigma_i \sigma_j/ \sigma_{\mathrm{eff}}$, where
$\sigma_i$, $\sigma_j$ and $\sigma_{\mathrm{ij}}$, are cross sections for single- and double-parton scatters of types $i$ and $j$.
Practically in all models it is assumed that transverse distribution of partons does not depend on $x$ of the parton\footnote{With an exception of the \pythia model
described in~\cite{Corke:2011yy}. However, this option is not used in the most recent \pythia tunes.}. 
In the approximation where the correlations between partons are neglected, the inclusive cross section of $N$ binary collisions is $\propto \sigma_{\mathrm{eff}}^{1-N}$. 
Hence the sensitivity to this  parameter should grow with the hadron multiplicity (usually characterized by charged-particle multiplicity in the experimental measurements).

It is interesting to note, that the transverse area in which partons are localized, as determined by the fits to data, are 
at least a factor of two smaller than indicated by the HERA data on hard exclusive processes. This suggest 
that one may need to include pQCD effects which lead to decrease of $\sigma_{\mathrm{eff}}$ with increase of the virtuality of the collision, see a review in~\cite{Blok:2017alw}. 
This pattern was implemented for example in~\cite{Blok:2015rka, Blok:2015afa}.

To extend studies of the low-$p_{\rm T}$ suppression mechanism we propose observables 
which minimize soft physics effects and still preserve sensitivity to the presence of the semi-hard collisions. 
We use here an observation that parton showers lead to a short-range correlation in rapidity, 
while a correlation of binary semi-hard collision extends to noticeably larger rapidity  intervals.
So we suggest to measure how the transverse momenta of hadrons produced in association with a trigger object are balanced 
as a function of rapidity. 
The exact definitions of the proposed observable is given  in Section~\ref{Sec:Observables}.
One of the advantages of such observable is that the contribution of the events where the trigger and the balancing particles belong to different parton-parton interactions should cancel, as long as the parton-parton interactions are independent.
This is in difference from the observables maximizing effects of MPI such as correlation of multiplicities at different rapidity intervals first considered by UA5 collaboration, see a review in \cite{Alner:1987wb}. 
These data were one of the first indications of the role of MPI in hadron-hadron collisions at  collider energies and the enhancement of MPI in the high multiplicity events.

Our numerical studies described below demonstrate sensitivity of the proposed variable to the assumed dynamics. 
A study of the same observable as a function of the multiplicity of final-state particles (which in the discussed models 
originate from fluctuation of the number of hard collisions or a combination of the soft and hard collisions)  
provides an additional discriminating tool which is a natural combination of the UA5-like and the inclusive transverse momentum balance observables. 
For high multiplicities the discussed observable is sensitive to effects such as screening or a formation of quark gluon plasma in collisions of protons. For these reasons we shall also study the observable as a function of the charged particle multiplicity.
Finally, we will investigate the impact of the so-called color-reconnection (CR) mechanism which is 
in continuous development by many Monte Carlo authors~\cite{Sjostrand:1993hi,Lonnblad:1995vr,Christiansen:2015yca,Argyropoulos:2014zoa,Cuautle:2016ukm,Bierlich:2015rha,Gieseke:2012ft,Gieseke:2017clv,Gieseke:2018gff}. 
The most of mechanisms discussed above are assumed to be dependent on the collision energy, therefore we shall perform our tests at two center-of-mass (CM) collision energies, $\sqrt{s}=$~7~TeV and $\sqrt{s}=$~13~TeV.

The paper is organized as follows. In Section \ref{Sec:Observables} we define the observables 
in a more formal way, while the justification of kinematic cuts is given in~Section~\ref{Sec:optimization}. 
In Section~\ref{Sec:MCmodels} a summary of the discussed models is presented.
The results of calculations using these models are presented in Section~\ref{Sec:Results}. 
Our conclusions are presented in Section~\ref{Sec:Conclusions}.

%% file: Observables.tex
As mentioned, we will be interested in a mechanisms of particle
production in hadron-hadron collisions, in particular in finding experimental
observables that are sensitive to a particular models. As is known,
the particle production is driven by the minijets, i.e. semi-hard
partons (quark and gluons) produced in a collision of incoming partons
(one or many), or in a bremsstrahlung process. 

Partons produced in  different mechanisms are, in general, correlated
in a different way. For example, if we concentrate on rapidity of
produced partons, we may expect that bremsstrahlung partons will have
short-range correlations, while the partons produced in a hard collision
will have a long range tails.

\begin{figure}
\begin{centering}
\flushleft A)
\par\end{centering}

\begin{centering}
\includegraphics[width=7cm]{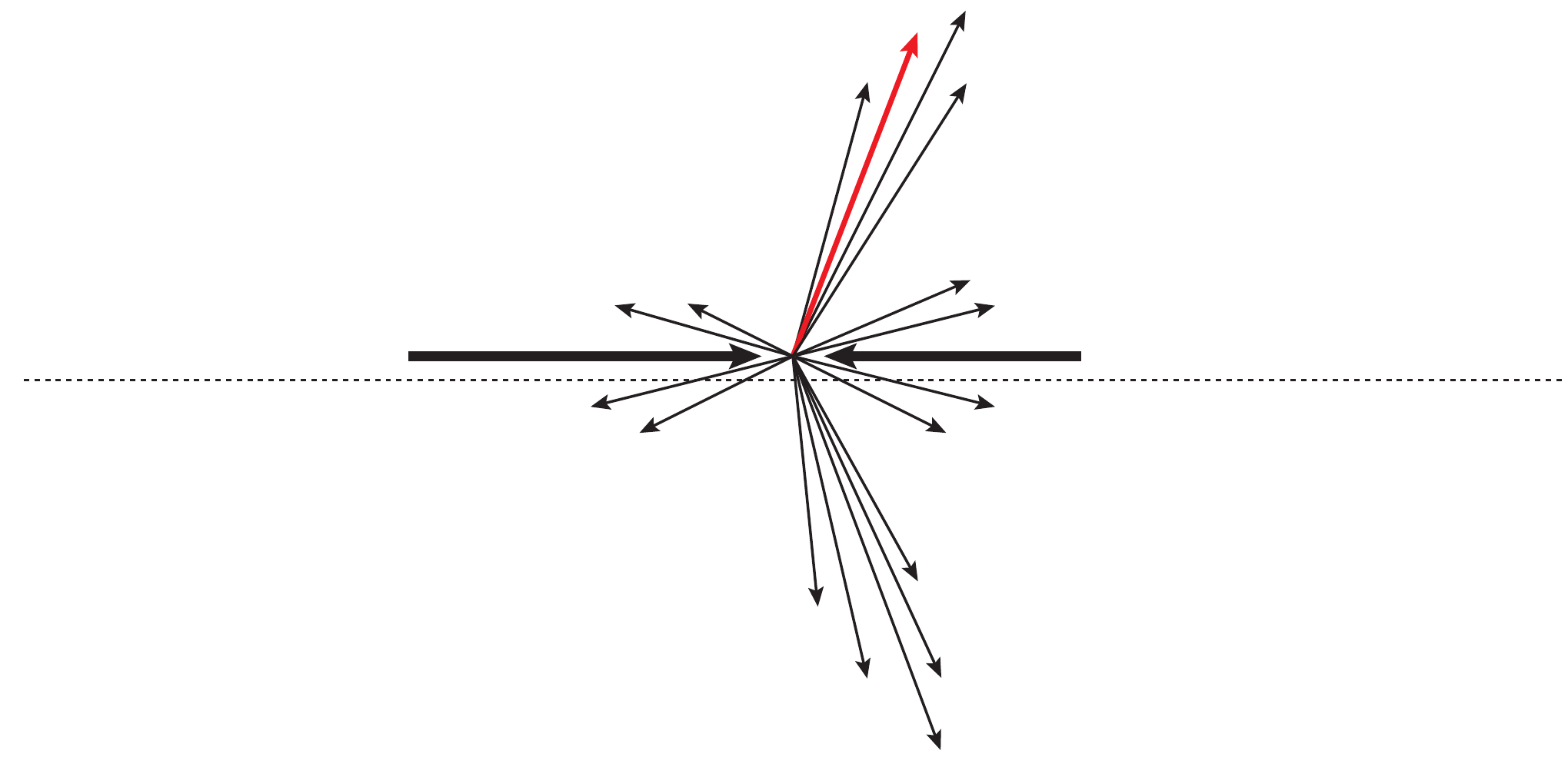}
\par\end{centering}

\begin{centering}
\flushleft B)
\par\end{centering}

\begin{centering}
\includegraphics[width=7cm]{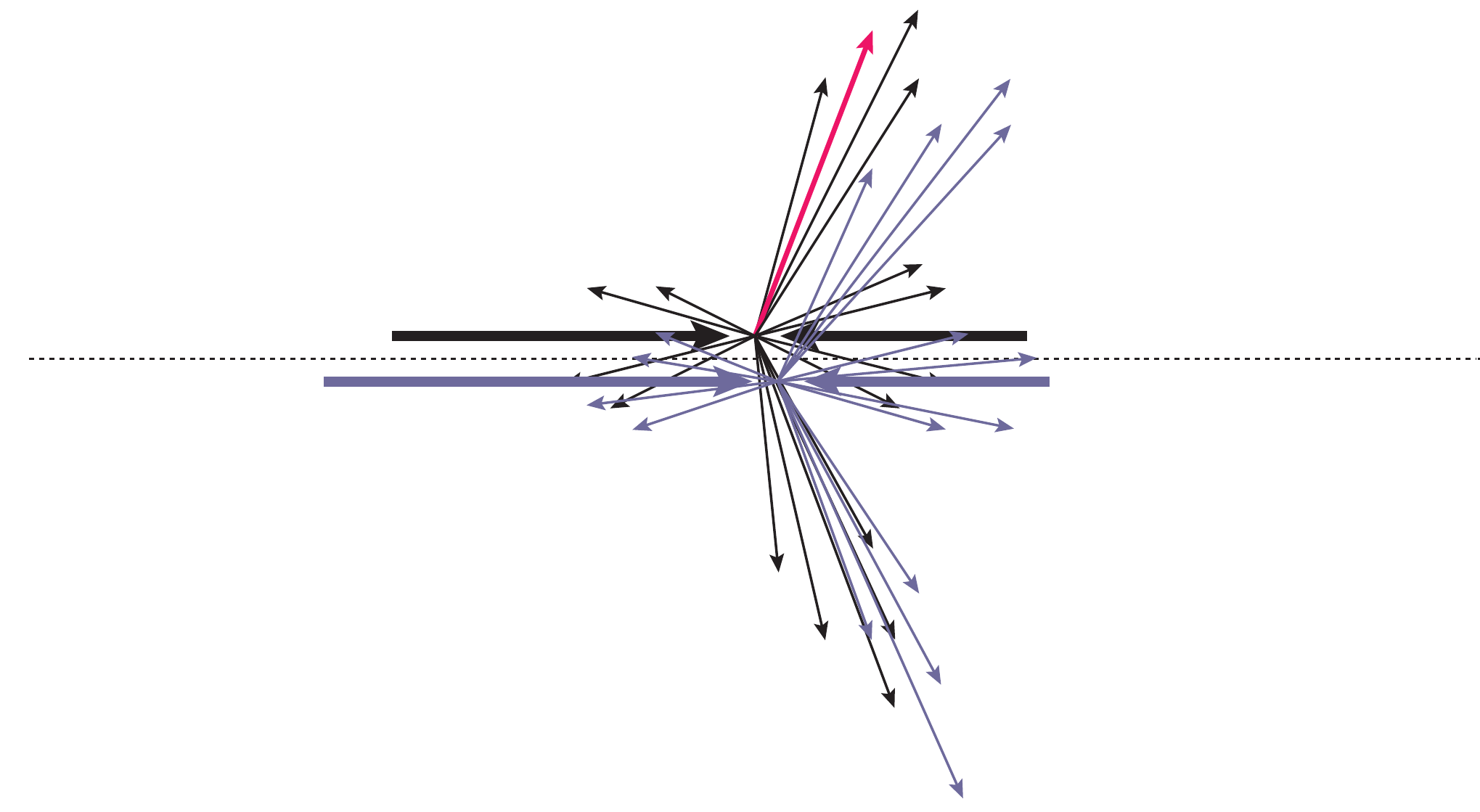}
\par\end{centering}

\caption{A) Single collision on the $y-z$ plane with partons produced due
to hard process and initial and final state radiation. The thick lines
represent the incoming partons while the red arrow represents the selected
trigger parton. The total transverse momentum of all partons sums up
to zero. B) An event with two hard collisions. For each hard collision
the momentum is conserved independently, if no correlations are present.
\label{fig:MPIcorr1}}
\end{figure}

One way to study the correlations is to investigate how the transverse
momentum is balanced as a function of rapidity. The practical observable
may be constructed as follows (see Fig.~\ref{fig:MPIcorr1}).
For a given event with $n$ final state particles, we pick up a particle
$k$ within a fixed rapidity interval and a certain (small) $p_{T}$.
Let us call this a trigger particle. Then, we define the total transverse
momentum of the all remaining final state particles along the trigger
particle, contained in a rapidity bin $\Delta\eta$:
\begin{equation}
\label{eq:obs}
p_{T}^{\mathrm{rec}\,\left(k\right)}\left(\eta\right)=\sum_{i=1,\dots n,\, i\neq k}
|\vec{p}_{Ti}|\cos \phi_i\,\, 
\Theta\left(\left(\eta-\frac{\Delta\eta}{2}\right)<\eta_{i}<\left(\eta+\frac{\Delta\eta}{2}\right)\right)\,,
\end{equation}
where $\Theta$ is the step function and $\phi_i$ is the azimuthal angle of the $i$-th particle, 
in the coordinate system where the $y$ axis is defined by the trigger particle $k$; 
in that system we simply add up the $y$ components of the recoil particles.
$p_{T}^{\mathrm{rec}\,\left(k\right)}\left(\eta\right)$ can be calculated
on the event-by-event basis so that we can define the average $\left\langle p_{T}^{\mathrm{rec}}\right\rangle \left(\eta\right)$
as
\begin{equation}
\left\langle p_{T}^{\mathrm{rec}}\right\rangle \left(\eta\right)=\frac{\sum_{k=1}^{N}p_{T}^{\mathrm{rec}\,\left(k\right)}\left(\eta\right)}{N}\,,
\end{equation}
where $N$ is the total number of events with the required trigger
particle present. 
We can also define similar quantity for the trigger particle, $\left\langle p_{T}^{\mathrm{trig}}\right\rangle \left(\eta\right)$,
by simply counting only the trigger particles.
The total momentum conservation requirement gives,
obviously,
\begin{equation}
\int d\eta\,\left\langle p_{T}^{\mathrm{rec}}\right\rangle \left(\eta\right)=\int d\eta\,\left\langle p_{T}^{\mathrm{trig}}\right\rangle \left(\eta\right)\,.
\end{equation}

%% file: KinematicCuts.tex
This section justifies the choice of final state objects used to study the mechanisms of the minijet production.
We are mostly guided by a  performance of the LHC general-purpose detectors, ATLAS and CMS.
Therefore, the usage of charged particles is the only option to study minijet production with upper 
$p_{\rm T}$ limit of a few GeV. 
The tracking system of the experiments allows to reliably reconstruct 
charged particles with $\eta <$~2.5(2.4) for ATLAS(CMS) starting from $p_{\rm T} \approx $ 250~MeV. 
Therefore, we chose $2.0<\eta<2.4$ for a trigger object in order to maximize the
possible $\eta$ distance for recoil particles. 

\begin{figure}[hbtp]
\begin{center}
\includegraphics[ width=0.5\textwidth]{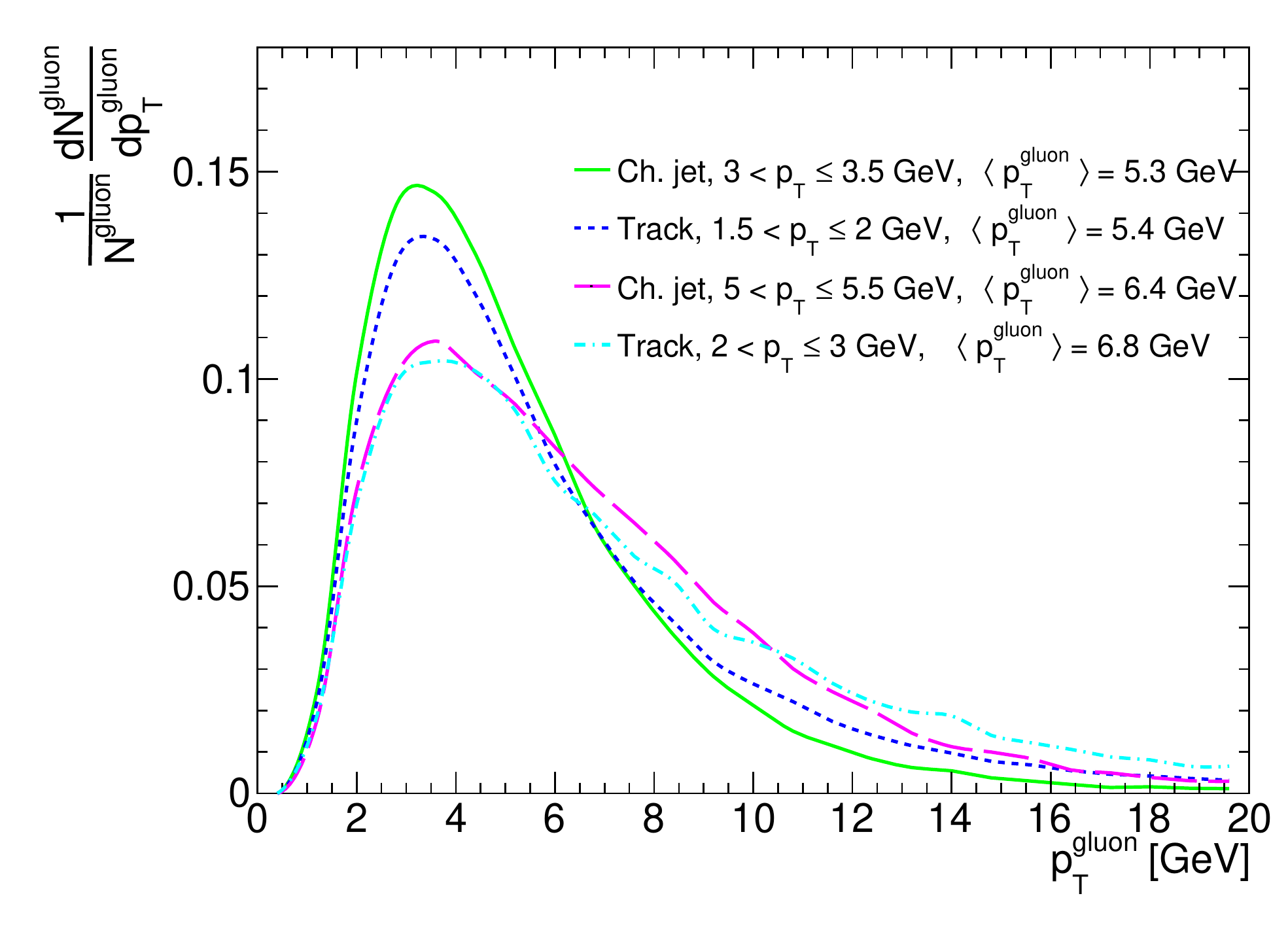}%
\caption{Gluon $p_{\rm T}$ distribution for various final-state triggers as obtained using \pythia 8 CUETP8M1 model.}%
\label{Trigger2GluonMatching}%
\end{center}
\end{figure}

There are two options for choosing the trigger object: (i) a single charged particle, or (ii)
 a charged-particle jet. Both approaches have their advantages. The single charged particle is a 
very simple and stable trigger, which is, in the contrast to the jet trigger, not contaminated by an additional activity from the UE.
The second option is expected to be better connected to the initial parton (mainly a gluon).
This is illustrated in Fig.~\ref{Trigger2GluonMatching} where  we investigate (with the help of \pythia) to
what $p_{\rm T}$ of initial gluon the final state trigger corresponds to. These distributions
are plotted under the assumption that the initial gluon, originating in the primary scattering or in MPI, 
can be matched with the final state trigger by a requirement of the maximum distance 
$R = \sqrt{(\phi_{\rm p}- \phi_{\rm t})^2+(\eta_{\rm p}- \eta_{\rm t})^2}$. Here 
 $\phi_{\rm p}$ ($\eta_{\rm p}$)  and $\phi_{\rm t}$ ($\eta_{\rm t}$) are azimuthal angles~(pseudorapidities)
of the initial gluon and the final state trigger object, respectively. 
We found that in \pythia~8 model   there is a strong spatial correlation between
the trigger objects and the parent gluons for the $p_{\rm T}$ range of interest. For $R<$~0.25 it is possible to match 80\% of them, thus that value is used to obtain
the distributions shown in Fig.~\ref{Trigger2GluonMatching}.  
The $p_{\rm T}$ windows of the trigger are chosen to be sensitive to the  suppression of the minijet production. 
One can see that the distributions are expectedly narrower for charged-particle jets than for single charged particle,
even if they correspond to the same gluon $\langle p_{\rm T} \rangle$. The distribution for single-particle trigger
has long tail that is quite noticeable for $p_{\rm T} > $~10~GeV. 
The main disadvantage of using the charged-particle jet is a contamination by UE. 
In order to reduce the UE contamination which grows with the jet area as $R^2$\cite{Dasgupta:2007wa} we use 
small distance parameter of $R=0.4$ in the anti-$k_{\rm T}$ jet clustering  algorithm~\cite{Cacciari:2008gp}.
In this case the UE contribution to the jet is $\sim$~0.5~GeV on average.

%% file: BasicMinijetMode.tex
As mentioned, the particle production mechanism is driven by $2\rightarrow2$
perturbative parton production.
 In the leading order (LO) the cross section for a production of two jets reads (only $gg\rightarrow gg$
channel is included here for simplicity):
\begin{multline}
\frac{d\sigma_{2\mathrm{jet}}}{dp_{T}^{2}dz_{1}dz_{2}}=\frac{1}{16\pi}\,\frac{1}{p_{T}^{4}}\,\frac{z_{1}\, z_{2}}{(z_{1}+z_{2})^{4}}\,\\
\, f_{g/H}\left(z_{1}+z_{2},\mu^{2}\right)f_{g/H}\left(\frac{p_{T}^{2}}{s}\,\frac{z_{1}+z_{2}}{z_{1}z_{2}},\mu^{2}\right)\,\frac{1}{2}\left|\overline{\mathcal{M}}\right|_{gg\rightarrow gg}^{2}\left(z_{1},z_{2}\right)\,,\label{eq:minijets1}
\end{multline}
where
\begin{equation}
\left|\overline{\mathcal{M}}\right|_{gg\rightarrow gg}^{2}\left(z_{1},z_{2}\right)=g^{4}\,\frac{9}{2}\,\frac{\left(z_{1}^{2}+z_{1}z_{2}+z_{2}^{2}\right)^{3}}{z_{1}^{2}z_{2}^{2}\left(z_{1}+z_{2}\right)^{2}}\,,
\end{equation}
is the LO matrix element squared and 
\begin{equation}
z_{1,2}=\frac{\left|\vec{p}_{T\,1,2}\right|}{\sqrt{s}}\, e^{\, y_{1,2}}.
\end{equation}
Above, $f_{g/H}$ are the gluon distributions in a hadron, $\mu^{2}$ is the hard scale $\sim p_{\rm T}^2$ , $s$ is the square of the CM energy, $\vec{p}_{T\, 1,2}$ are the transverse momenta of the outgoing partons while $y_{1,2}$ are their rapidities. 
Due to the momentum conservation we have at LO $\left|\vec{p}_{T1}\right|=\left|\vec{p}_{T2}\right|\equiv p_{T}$.

 There are two related aspects of this
mechanism which are relevant at small transverse momenta \cite{Sjostrand1987}.
First, the dijet cross section is divergent for jet $p_{T}\rightarrow0$:
\begin{equation}
\frac{d\sigma_{2\mathrm{jet}}}{dp_{T}^{2}}\sim\frac{\alpha_{s}^{2}\left(p_{T}^{2}\right)}{p_{T}^{4}}\,.
\end{equation}
It is however expected that the growth of the spectrum is tamed by some mechanism
already in the perturbative domain for $p_{T}\sim2-3\,\mathrm{GeV}$.
In phenomenological model of \cite{Sjostrand1987} the suppression
factor was introduced as follows:
\begin{equation}
\frac{d\sigma'_{2\mathrm{jet}}}{dp_{T}^{2}}=\frac{d\sigma_{2\mathrm{jet}}}{dp_{T}^{2}}\,\frac{p_{T}^{4}}{\left(p_{T}^{2}+p_{T0}^{2}\left(s\right)\right)^{2}}\,\frac{\alpha_{s}^{2}\left(p_{T}^{2}+p_{T0}^{2}\left(s\right)\right)}{\alpha_{s}^{2}\left(p_{T}^{2}\right)}\,,\label{eq:MinijetReg-1}
\end{equation}
where $p_{T0}\left(s\right)$ is a cutoff parameter which depends
on the total CM energy of the collision $s$
\begin{equation}
\label{eq:ptevol}
p_{T0}(s) = p_{T0}^{\mathrm{ref}} \left(\frac{s}{s_0}\right)^{\lambda}\, ,
\end{equation}
where $p_{T0}^{\mathrm{ref}}$, $s_0$ and $\lambda$ are parameters to be determined from the data.

Second, even with the cutoff, the inclusive dijet cross section can exceed the total inelastic 
cross section, implying presence of events with multiple hard parton-parton collisions. The average number of the  parton collisions is defined to be $\left\langle n\right\rangle \sim\sigma_{2\mathrm{jet}}/\sigma_{\mathrm{ND}}$,
where $\sigma_{\mathrm{ND}}$ is the nondiffractive total cross section, since the production of jets in diffraction is strongly suppressed.

\begin{figure}
\begin{centering}
\includegraphics[width=8cm]{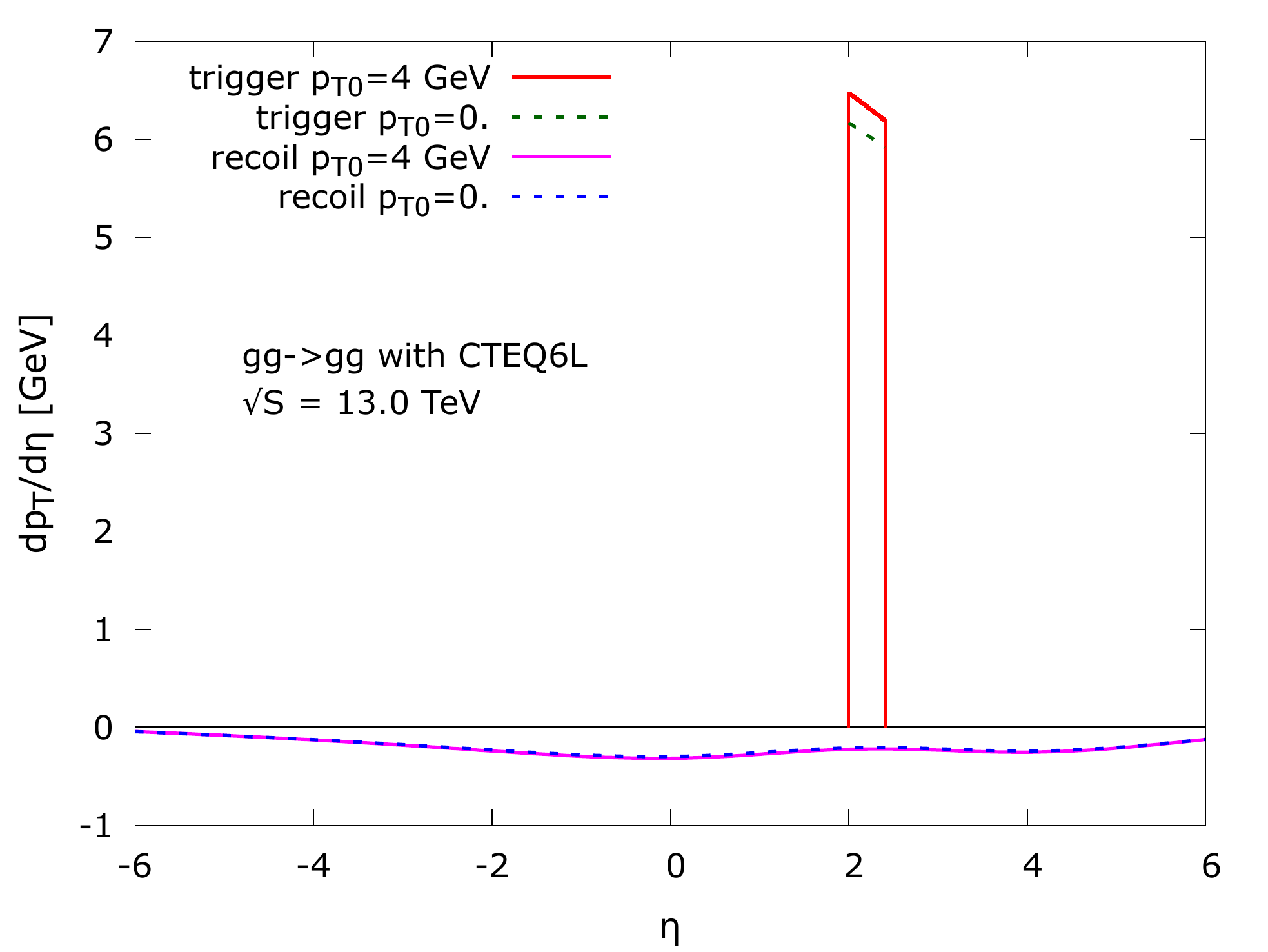}\quad \includegraphics[width=8cm]{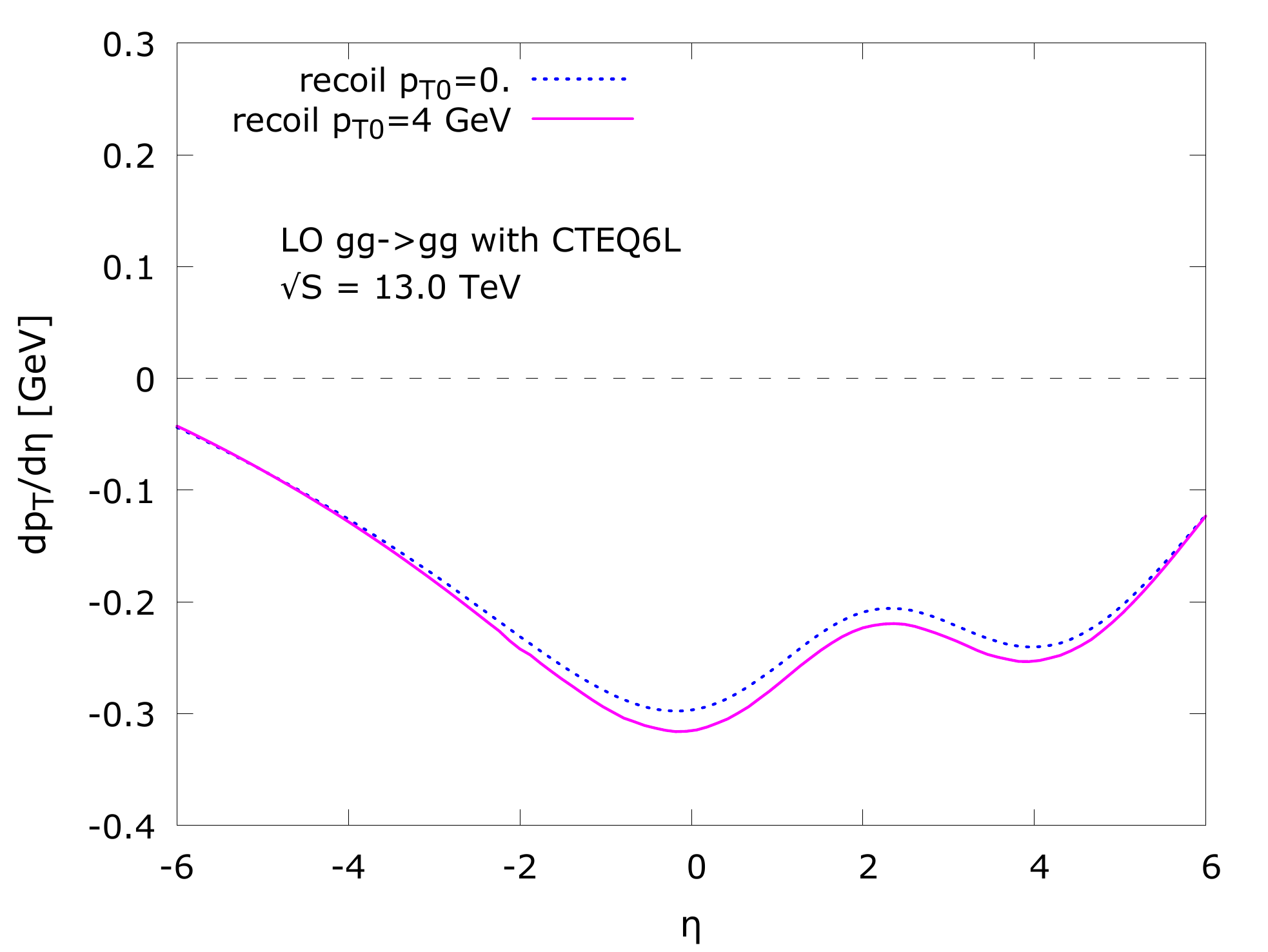}
\par\end{centering}

\caption{Left: Rapidity correlations $\left\langle p_{T}\right\rangle _{\mathrm{trig}}$ and $\left\langle p_{T}\right\rangle _{\mathrm{rec}}$
from the basic QCD perturbative model with the $p_T$ cutoff. The trigger has $2.0<p_T<3.0\,\mathrm{GeV}$ and rapidity $2.0<y<2.4$. Right: Zoom of the recoil system curves. \label{fig:pTmean}}

\end{figure}

In the above simple model we can simply obtain the expressions for our main observable, i.e. the average
transverse momenta of the trigger and the recoil system: $\left\langle p_{T}^{\mathrm{trig}}\right\rangle \left(y\right)$ and $\left\langle p_{T}^{\mathrm{rec}}\right\rangle \left(y\right)$, respectively. 
 Using the kinematic cuts discussed in Section~\ref{Sec:optimization}, we show the sample result in Fig.~\ref{fig:pTmean}. In this case, the trigger has $2.0<p_T<3.0\,\mathrm{GeV}$ and rapidity $2.0<y<2.4$. The result shows a typical pattern of the rapidity correlations encoded in the hard matrix element and will be a useful reference point when discussing realistic mechanisms.

%% file: PythiaModel.tex
A detailed description of the model is contained in the series of
papers \cite{Sjostrand1987,Sjstrand2004,Sjostrand2005,Corke2011a}.
Here we give only a very brief summary.

The essential point is that the hard collisions are not completely
independent. This is true in several respects. First, the hard collisions contributing to a single event
are ordered according to a scale $\sim$ $p_{T}$ and interleaved with the shower mechanisms. Second, the correlations
are introduced by the proper treatment of the beam remnants. That
is, the removal of a parton from the beam affects the remaining multi-parton
distribution function in the longitudinal and flavour space. Finally,
there are correlations in the transverse momentum space introduced
by the so-called primordial $k_{T}$. Also, the colour reconnection introduces correlations.

The event generation goes as follows. After
generating a hard interaction with certain $p_{T\,\mathrm{max}}$,
the following step, i.e. an emission with transverse momentum $p_{T}<p_{T\,\mathrm{max}}$,
is described by the probability distribution
\begin{multline}
\frac{dP}{dp_{T}}=\left(\frac{dP_{\,\mathrm{MPI}}}{dp_{T}}+\sum\frac{dP_{\,\mathrm{IS}}}{dp_{T}}+\sum\frac{dP_{\,\mathrm{FS}}}{dp_{T}}\right)\\
\times\exp\left\{ -\int_{p_{T}}^{p_{T\,\mathrm{max}}}dp'_{T}\left(\frac{dP_{\,\mathrm{MPI}}}{dp'_{T}}+\sum\frac{dP_{\,\mathrm{IS}}}{dp'_{T}}+\sum\frac{dP_{\,\mathrm{FS}}}{dp'_{T}}\right)\right\} \,,\label{eq:Pythia_evol}
\end{multline}
where the subsequent probabilities in brackets correspond, respectively, 
to the probability distribution of another hard collision, the emission
from the initial state, and the final state emission. The exponential
`Sudakov form factor' originates from the requirement that no emission
took place between $p_{T}$ and $p_{T\,\mathrm{max}}$. The initial
and final state showers are based on the DGLAP evolution and we do
not discuss them here. The MPI probability distribution is 
impact parameter dependent. For the hardest event it reads 
\begin{equation}
\frac{dP_{\mathrm{MPI}}}{dp_{T}d^{2}b}=\frac{\mathcal{O}\left(b\right)}{\left\langle \mathcal{O}\right\rangle }\,\frac{1}{\sigma_{\mathrm{nd}}}\,\frac{d\sigma}{dp_{T}}\,\exp\left\{ -\int_{p_{T}}^{p_{T\,\mathrm{max}}}dp'_{T}\,\frac{\mathcal{O}\left(b\right)}{\left\langle \mathcal{O}\right\rangle }\,\frac{1}{\sigma_{\mathrm{nd}}}\,\frac{d\sigma}{dp'_{T}}\right\} \,,\label{eq:MPI_probab}
\end{equation}
where the cross section $d\sigma/dp_{T}$ is given by the basic minijet
model, Eq.~(\ref{eq:MinijetReg-1}). The matter overlap function $\mathcal{O}\left(b\right)$
is
\begin{equation}
\mathcal{O}\left(b\right)\propto\int dt\,\int d^{3}r\,\rho\left(x,y,z\right)\rho\left(x+b,y,z+t\right)\,,
\label{eq:Overlap_Func}
\end{equation}
were $\rho$ is the matter distribution in a single hadron. In the recent Pythia
version the default setting is that the overlap function $\mathcal{O}\left(b\right)$ is of the form $\exp\left(-b^{\mathrm{Pow}}\right)$, where $\mathrm{Pow}=1.85$. By default, there is  no $x$ dependence in the matter distribution. However, it
is possible to choose a non-standard setting where the width of the 
Gaussian matter distribution depends on $x$ as $1 + a_1 \log (1 / x)$.
The average $\left\langle \mathcal{O}\right\rangle $ is
defined in a special way taking into account that every event has
to have at least one collision, see the original papers for details.

The actual number $n$ of binary collisions is determined by the truncation
of the iterative procedure when no further emissions can be resolved
from Eq.~(\ref{eq:Pythia_evol}).

The effect of the interleaved evolution (\ref{eq:Pythia_evol}) is
most important for the initial state shower and MPI. This is because
they compete for the beam energy. The actual correlations are incorporated
by modifying the beam remnant in regards to the remaining longitudinal
momentum and flavour.

We have studied the observables proposed in Section~\ref{Sec:Observables} in 
the context of the dependence on various parameters, notably on the $p_{\rm{T}0}$. 
These exercises aim mostly at understanding some important aspects of the event generator, 
and are not meant to be compared with any discriminating data.
Therefore, we include these studies in Appendix~\ref{Sec:Correlations} as a useful reference to 
various correlation effects. In fact changing only for example
$p_{\rm{T}0}$ modifies the tune and the description of certain data will be spoiled, therefore
our main results are provided for unchanged tunes of Pythia 8 that are meant to describe UE, Minimum Bias (MB), and DPS data.
Namely, we choose tunes constructed with leading and next-to-next-to-leading order (NNLO) PDF sets:
CUETP8M1~\cite{Khachatryan:2015pea}, and very recent CMS tunes: CP2, CP4, CP5~\cite{CMS:2018zub}. 
%
Selected tunes describe MB and UE data at a similar level~\cite{Khachatryan:2015pea,CMS:2018zub}, however they have a significantly different values 
of the parameters.
The first and second tunes are based on LO PDF sets, the third and  forth were NNLO ones. 
The difference between PDF sets is reflected in the $p_{\rm T}$
distributions of gluons coming from primary and MPI interactions (see curves for CUETP8M1 and CP5 tunes in Fig.~\ref{GluonPtDistribution}).
The other key feature is a choice of  the impact-parameter distribution. For CUETP8M1 tune an exponential 
overlap function is used, while for the new tunes (CP2, CP4, CP5) a double-Gaussian matter distribution function 
(see Eq.~(\ref{eq:Overlap_Func})). For given choice of PDF sets and impact-parameter profiles 
a number of parameters is tuned, including parameters of the $b$-profile, smooth cutoff parameter $p_{\rm T0}$, 
colour reconnection ones, and some others. It is worth noting  that the tunes differ in simulation 
of the initial state radiation (ISR). \pythia 8 CUETP8M1  and CP5 have rapidity ordering  for ISR,
while it is switched off for CP2 and CP4 tunes. It is shown in Section~\ref{Sec:Results} that  the mechanism has an impact on the
the studied rapidity correlation.

%% file: HerwigMPImodel.tex
The MPI model used in \herwig{} has been reviewed several 
times~\cite{Butterworth:1996zw,Borozan:2002fk,Bahr:2008dy,Bahr:2008pv}. 
Here we aim just to describe briefly the most important building blocks and the parameters of the model.

\begin{figure*}[]
\begin{center}
  \includegraphics[width=0.49\textwidth]{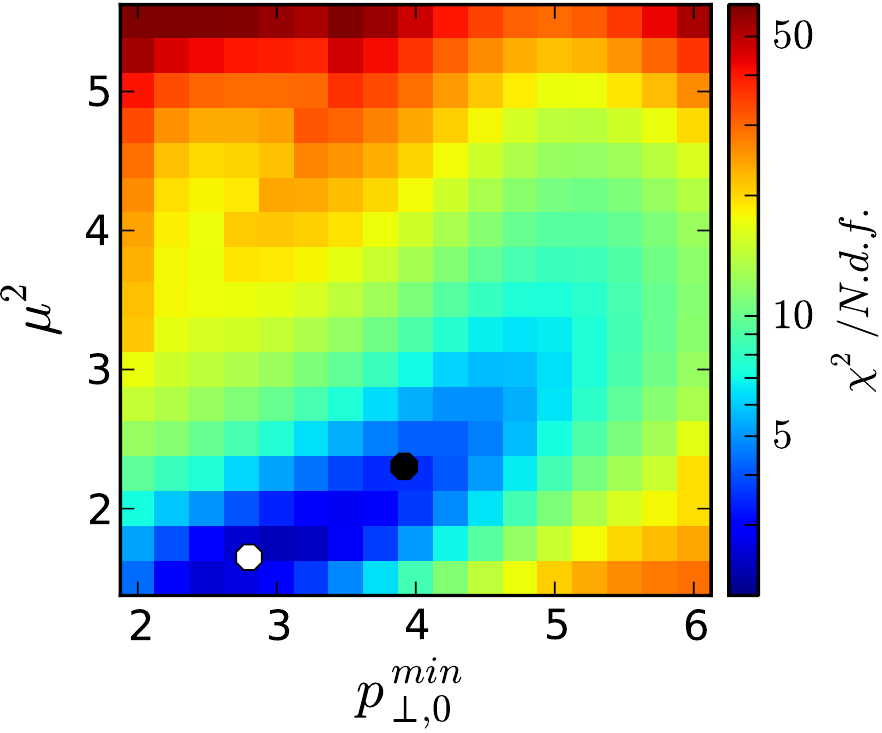}
  \includegraphics[width=0.49\textwidth]{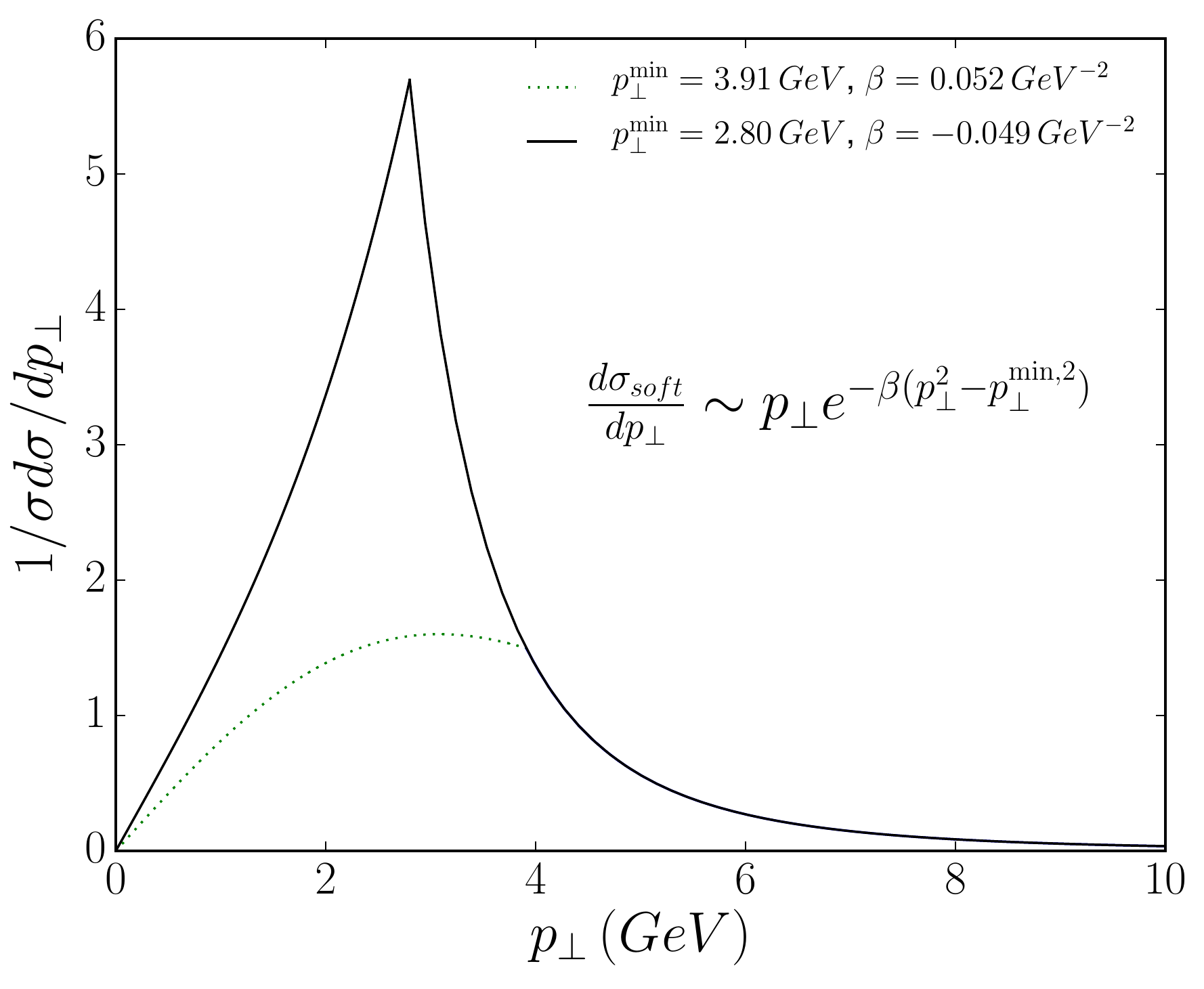}
  \end{center}
  \caption{Left panel: The weighted $\chi^2/N\!.d.f.$ as a function of $\mu^2$ and
     $\ptminnought$ from~\cite{Seymour:2013qka}. The best fit point is shown with a white dot (tune Var2) 
     and a black dot (tune Var1) represents 
     a good fit with higher $\ptminnought$ which is preferred by the CDF \seff data~\cite{Bahr:2013gkj}. 
     Right panel: an extension of $p_{\perp}$ into non-perturbative regime with
``Gaussian'' transverse momentum distribution for two parameters sets denoted by circles on the plot in the left panel.
}
  \label{fig:ptmin}
\end{figure*}

The model is formulated in the impact parameter space. At a fixed impact
parameter, multiple parton scatterings are assumed to be independent,
however, later they are correlated, for example, by imposing  energy-momentum conservation 
or through the colour reconnection mechanism. 
There are two types of parton-parton scatterings in the model, soft and semi-hard, the both are separated 
by a transverse momentum scale \ptmin, which is one of the main tuning parameters in the model. 
The value of \ptmin is allowed to vary with energy and the evolution is govern by a power law, see Eq.~(\ref{eq:ptevol}).   
In fact, it is $\ptminnought=\ptmin(7~\textrm{TeV})$ and power $\lambda$ that is fitted to data.
Below \ptmin, scatters are assumed to be non-perturbative, with valence-like longitudinal momentum distribution and
``Gaussian'' transverse momentum distribution, see right panel of Fig.~\ref{fig:ptmin} for two examples
how the extrapolation to non-perturbative region can be realized in the model. 
Above
\ptmin, scatters are assumed to be perturbative, and take place
according to leading order QCD matrix elements convoluted with inclusive
PDFs and an overlap function~$A(b)$:

\begin{equation}
  A(b) = \int \mathrm{d}^2b_1 \, G(b_1) \int \mathrm{d}^2b_2 \, G(b_2)
  \, \delta^2(\mathbf{b}-\mathbf{b}_1+\mathbf{b}_2) \, ,
\end{equation}
where 
\begin{equation}
  G(b) = \frac{\mu^2}{4\pi}(\mu b)K_1(\mu b)
\end{equation}
is Fourier transform of dipole form factor $\frac{1}{(1-t/\mu^2)^2}$ which leads to overlap function
\begin{equation}
\label{eq:overlap}
  A(b) = \frac{\mu^2}{96\pi}(\mu b)^3K_3(\mu b),
\end{equation}
both $G(b)$, and $A(b)$ are normalised to unity. $K_i(x)$ is the modified Bessel function of the $i$-th kind and $\mu$ 
is the another important parameter of the model which governs the transverse distribution of partons in the proton and 
can be interpreted as an effective inverse proton radius. 
A lower values of $\mu$ lead to the broader 
matter distribution, therefore higher probability of peripheral collisions. 
Soft scatters might see a different matter distribution, therefore 
the model allows them to have a different inverse radius $\mu_{\rm{soft}}$ but it keeps the 
functional form of the overlap function from Eq.~\ref{eq:overlap}.
The two soft MPI parameters $\mu_{\rm{soft}}$ and $\sigma_{\rm{soft}}$, the non-perturbative cross section below \ptmin, 
are fixed by the inelastic hadron-hadron cross section and the b-inelastic slope parameter,
therefore they are not free parameters of the model.

The probability distribution of number of scatters is Poissonian at a
given impact parameter, but the distribution over impact
parameter is a considerably broader than Poissonian. The number
of soft and hard scatters is chosen according to this distribution and
generated according to their respective distributions. Each hard scatter
is evolved back to the incoming hadron according to the standard parton
shower algorithm, therefore the evolution is not interleaved like in Pythia model and additional 
scatterings are not ordered in any kinematic variable. Energy-momentum conservation is imposed 
by rejecting any scatters that take the total energy extracted from the
hadron above its total energy. 
As mentioned before the individual scatters might be colour correlated 
using a~colour reconnection model, described in detail in Ref.~\cite{Gieseke:2012ft}, in that 
model a reconnection probability \preco is applied. To summarize there are four 
main parameters of the model: \ptminnought\  
$\Lambda$, $\mu^2$ and \preco, which are fitted to the experimental data.
Unlike Pythia, Herwig does not have a large family of tunes, usually no more than one tune 
is released with a new version of the program. Therefore, in order to study in a meaningful way
effect of the parameters variation in \herwig we decided to use the 
two tunes prepared for the same version of the program~\cite{Seymour:2013qka}.  
The both tunes, which we label by Var1 (the default tune of \herwig{++ 2.7}) and Var2,
provide  good description
of the UE data over the collision energy range from 300 GeV to 7 TeV. This is visualized 
in the left panel of Fig.~\ref{fig:ptmin} (see~\cite{Seymour:2013qka} for the details), where we show 
the $\chi^2/N\!.d.f.$ value of the fit as a function of $\ptminnought$ and $\mu^2$, the both tunes 
are marked by black (Var1) and white (Var2) dots. 
A visible strong correlation between $\ptminnought$ and~$\mu^2$ (a long thin blue valley in Fig.~\ref{fig:ptmin})
reflects the fact that a smaller hadron radius means more likely central collisions 
and as a consequence more multiple scattering, which can be compensated to give a similar amount of
underlying-event activity by having fewer perturbative MPIs, i.e.~a
larger value of $\ptmin$.
The best fit value is for $\ptminnought={2.80}\,\mathrm{GeV}$ and
$\mu^2={1.65}\,\mathrm{GeV}^2$ (tune Var2), but one can obtain good fits 
for higher value of $\ptminnought={3.91}\,\mathrm{GeV}$, together with
$\mu^2={2.3}\,\mathrm{GeV}^2$ (tune Var1). As one can see from Fig.~\ref{fig:ptmin}~(right-panel) 
the $p_{\perp}$ spectra looks significantly different for these two tunes~\footnote{For all 
other parameters of the tunes see the Appendix.}, therefore they are well suited for our studies. 
It is worth to mention the default \herwig{++ 2.7} tune
gives the value of $\sigma_{\mathrm{eff}}=14.8$ mb which is 
close to $\sigma_{\mathrm{eff}}$ obtained from the combination of 
the two most precise experimental results for this observable from CDF and D0 measurements  
$\sigma_{\mathrm{eff}} = (13.9\pm1.5)~\mbox{\mbox{mb}}$\footnote{
The most of the $\sigma_{\mathrm{eff}}$ agrees with this value, however
it is worth noting that for example analysis of the exclusive photoproduction of $J/\psi$~\cite{Frankfurt:2010ea} 
suggests larger values of $\sigma_{\mathrm{eff}}$.}.

Finally, we will also show results of \herwig 7 which has new model for soft interactions
including diffractive final states and multiple particle production in multiperipheral kinematics, 
see~\cite{Gieseke:2016fpz} for the details.

%% file: Sherpa.tex
Minimum bias events in Sherpa~\cite{Gleisberg:2008ta} 
are simulated using the Shrimps package~\cite{Sherpawww,Schulz:2016vml}
which is based on Khoze-Martin-Ryskin (KMR) model~\cite{Ryskin:2009tj}. 
The KMR model is 
a multi-channel eikonal model in which the incoming hadrons are described as a 
superposition of Good-Walker states, which are diffractive eigenstates that diagonalize 
the T-matrix. Each combination 
of colliding Good-Walker states gives rise to a single-channel eikonal. The final eikonal 
is the superposition of the single-channel eikonals. The number of Good-Walker states is 
two in Shrimps (the original KMR model includes three states).
Each single-channel eikonal can be seen as the product of two parton densities, 
one from each of the colliding Good-Walker states. The evolution of the parton 
densities in rapidity due to extra emissions and absorption on either of the 
two hadrons is described by a set of coupled differential equations. 
The parameter $\Delta$, which can be interpreted as the Pomeron intercept, 
is the probability for emitting an extra parton per unit of rapidity. 
The strength of absorptive corrections is quantified by the parameter $\Lambda$,
which can be related to the triple-Pomeron coupling. A small region of size 
$\Delta Y$ around the beams is excluded from the evolution due to the finite 
longitudinal size of the parton densities. The boundary conditions for the parton 
densities are form factors, which have a dipole form characterized by the 
parameters $\Lambda^{2}$, $\beta_0^2$, $\kappa$ and $\xi$.
In this framework the eikonals and the cross sections for the different modes 
(elastic, inelastic, single- and double-diffractive) are calculated. 

Inelastic events are generated by explicitly simulating the exchange and re-scattering 
of gluon ladders. The number of primary ladders is given by a Poisson distribution whose 
parameter is the single-channel eikonal. The decomposition of the incoming hadrons into 
partons proceeds via suitably infrared continued PDFs.

The emissions from the ladders are then generated in a Markov chain. The pseudo-Sudakov 
form factor contains several factors: an ordinary gluon emission term, a factor accounting 
for the Reggeisation of the gluons and a recombination weight taking absorptive corrections 
into account. The emission term has the perturbative form $\alpha_s(k_T^2)/k_T^2$, 
that, as we have already seen in \pythia{} and \herwig{} models needs to be continued into the infrared region. 
In \sherpa{} in the case of $\alpha_s$ the transition 
into the infrared region happens at $Q_{\rm as}^2$ while in the case of $1/k_T^2$ the transition 
scale is generated dynamically and depends on the parton densities and is scaled by $Q_0^2$.

The propagators of the filled ladder can be either in a colour singlet or octet state, 
the probabilities are again given through the parton densities. The probability for a 
singlet can also be regulated by hand through the parameter $\chi_S$. A singlet propagator 
is the result of an implicit re-scattering.

After all emissions have been generated and the colours assigned, further radiation 
is generated by the parton shower. The strength of 
radiation from the parton shower can be regulated with $K_{\rm T}^{2}\_Factor$, which multiplies 
the shower starting scale. After parton showering partons emitted from the ladder 
or the parton shower are subject to explicit re-scattering, i.e. they can exchange 
secondary ladders. The probability for the exchange of a re-scattering ladder is 
characterised by $RescProb$. The probability for re-scattering over a singlet propagator 
receives an extra factor $RescProb1$.
After all ladder exchanges and re-scatterings, the colour can be rearranged in the event in a 
similar fashion to the colour reconnection models in \pythia{} and \herwig{}. 
Finally, the event is hadronized using the standard Sherpa cluster hadronization. 
In our studies we used the default settings of Multiple Interaction Models in Sherpa 2.2.2,
which is the only existing tune of the Shrimps model, for completeness we list its parameters in the Appendix~\ref{Shepa-app}.

%% file: Results.tex
In this Section we present the calculation of the observables defined in Section~\ref{Sec:Observables} using a few recent versions and tunes of \pythia, \herwig, and  \sherpa.  
Since some model parameters depend on $\sqrt{s}$, including the suppression of jet cross-section 
at very low $p_{\rm T}$, the results are presented for the CM energy $7$ and $13$ TeV. For the analysis we use non-diffractive 
inelastic events. The two approaches to the trigger object are studied, as discussed in Section~\ref{Sec:optimization}, first using a single charged particle as a trigger, 
second using a charged-particle jet as a trigger. The former is more model dependent, but is less affected by the UE contribution, while the latter provides a better connection to the parent parton. 

\begin{figure}[]

\begin{minipage}[h]{0.49\linewidth}
\center{\includegraphics[width=1\linewidth]{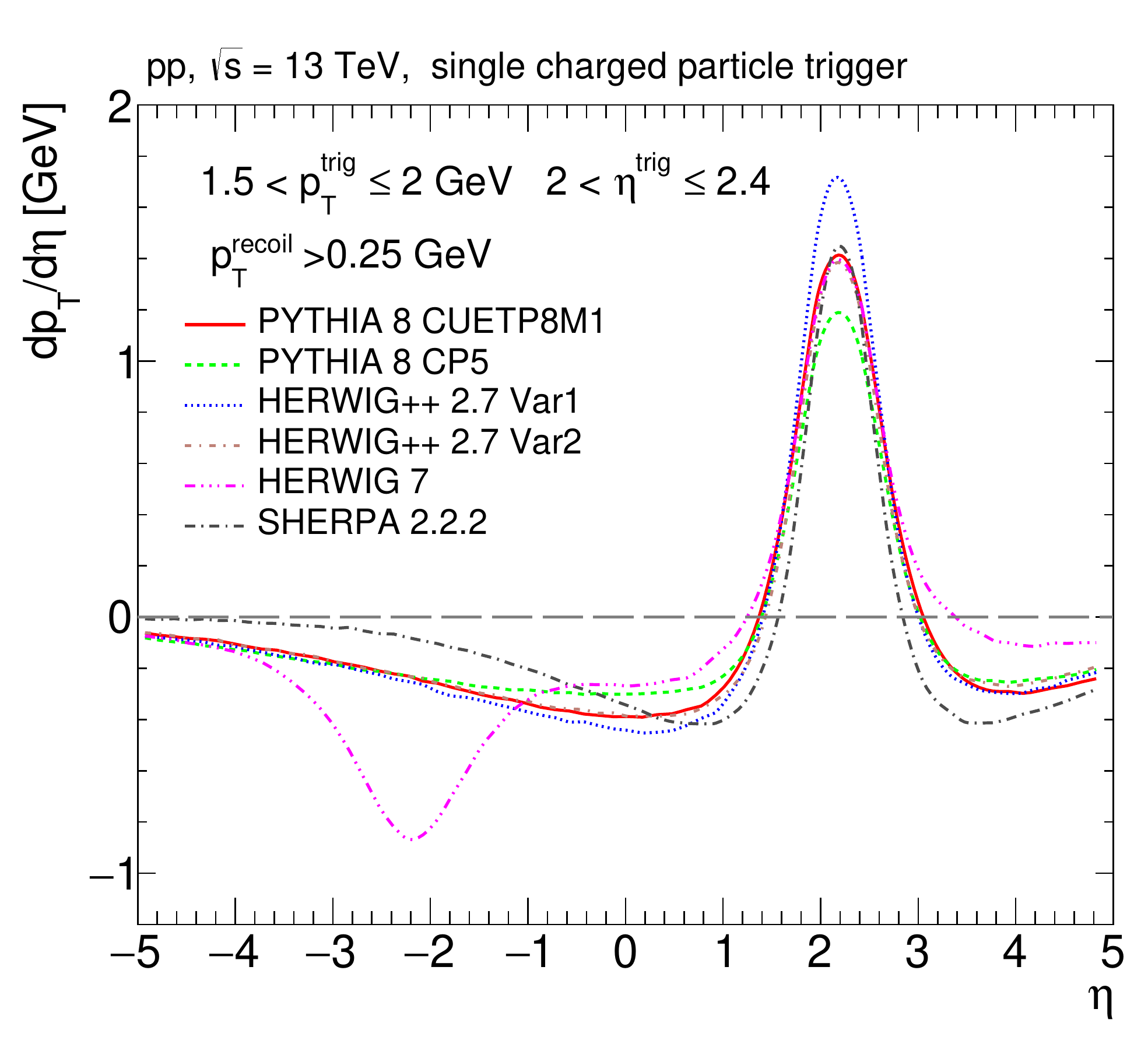} \\ (a)}
\end{minipage}
\begin{minipage}[h]{0.49\linewidth}
\center{\includegraphics[width=1\linewidth]{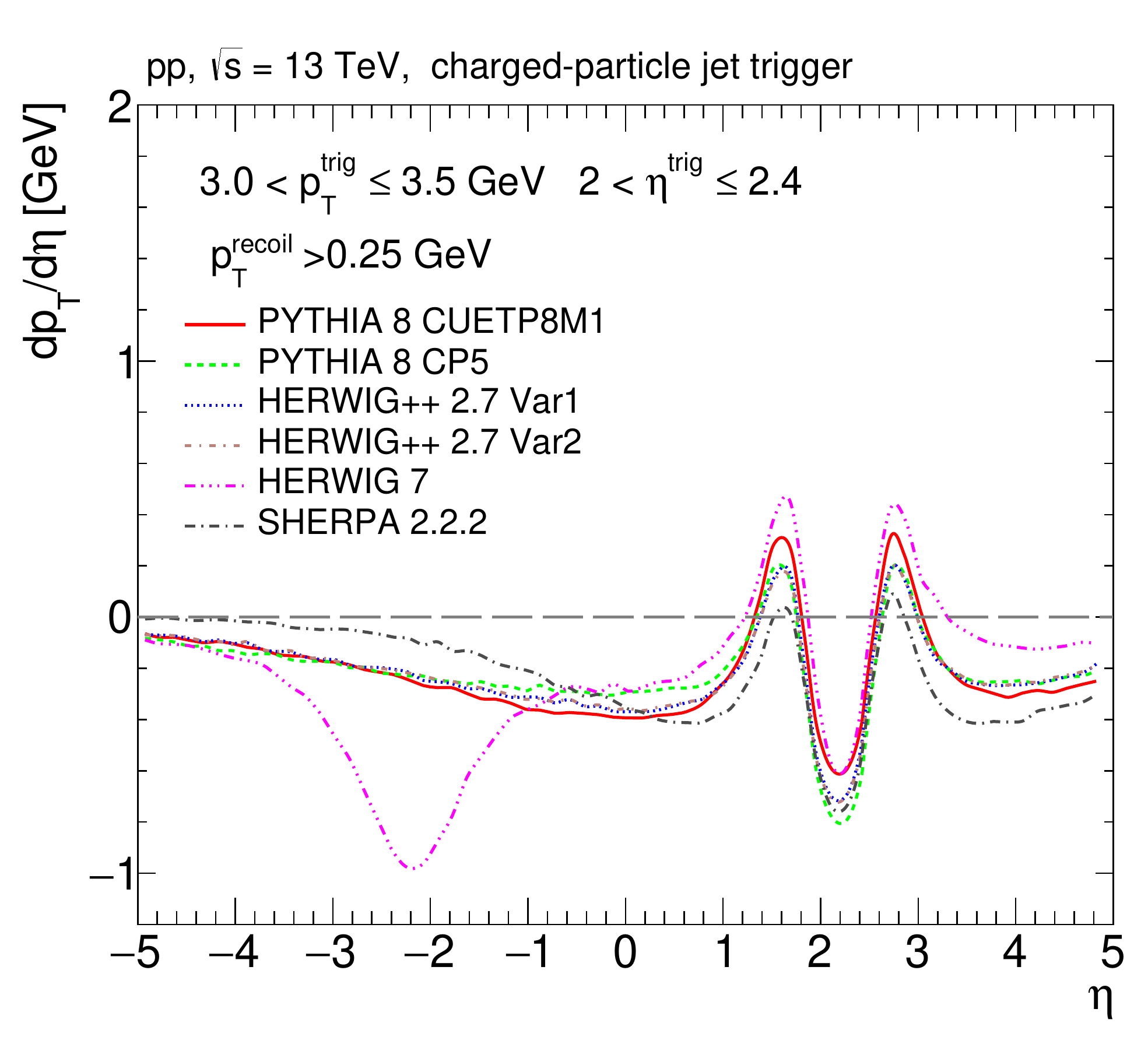} \\ (b)}
\end{minipage}

\caption{Rapidity correlation of recoiled system with respect to single-charged particle 
with $1.5 \leq p_{\rm T}< 2.0$ GeV (a) and with respect to charged-particle jet
with $3.0 \leq p_{\rm T}< 3.5$ GeV (b) at $\sqrt{s}=13$~TeV.}%
\label{RapidityCorrTrack13TeV}
\end{figure}

Fig.~\ref{RapidityCorrTrack13TeV} shows $p_{\rm T}^{\rm{rec}}$ as a function of pseudorapidity 
at $\sqrt{s}=13$~TeV  for the two trigger approaches. In the trigger region (near $2.0< \eta <2.4$), 
in the case of charged particle trigger (left panel),
we see a peak that is mostly caused by particles strongly correlated to a triggered particle (which would be clustered to the same jet).
In the case of the jet trigger technique, the particles clustered into the trigger jet 
are excluded from the calculation of $p_{\rm T}^{\rm rec}$, see Eq.~(\ref{eq:obs}), therefore we see a dale.
In the region distant from the trigger object ($\eta<$~0), one can see that
the difference between the trigger approaches is rather modest in the \pythia model.
This is expected, since the $p_{\rm T}$ ranges for the two trigger objects correspond,  on average, to the same 
$p_{\rm T}$ of the parent gluon (see Fig.~\ref{GluonPtDistribution}). 
The difference between the trigger techniques  is  stronger  for the other  models, i.e. \herwig{} and \sherpa.
This is probably due to different fragmentation and hadronization models. 
A peculiar feature of \herwig{ 7} is that it generates a recoil peak that lies at opposite $\eta$ region 
with respect to the trigger object, this seems to be a feature of new Soft MPI model~\cite{Gieseke:2016fpz}, which we plan 
to investigate more in the future. Finally, it is worth noticing that \sherpa produces the least amount of long-range correlations.

\begin{figure}[hbtp]
\begin{center}
\includegraphics[ width=1.0\textwidth]{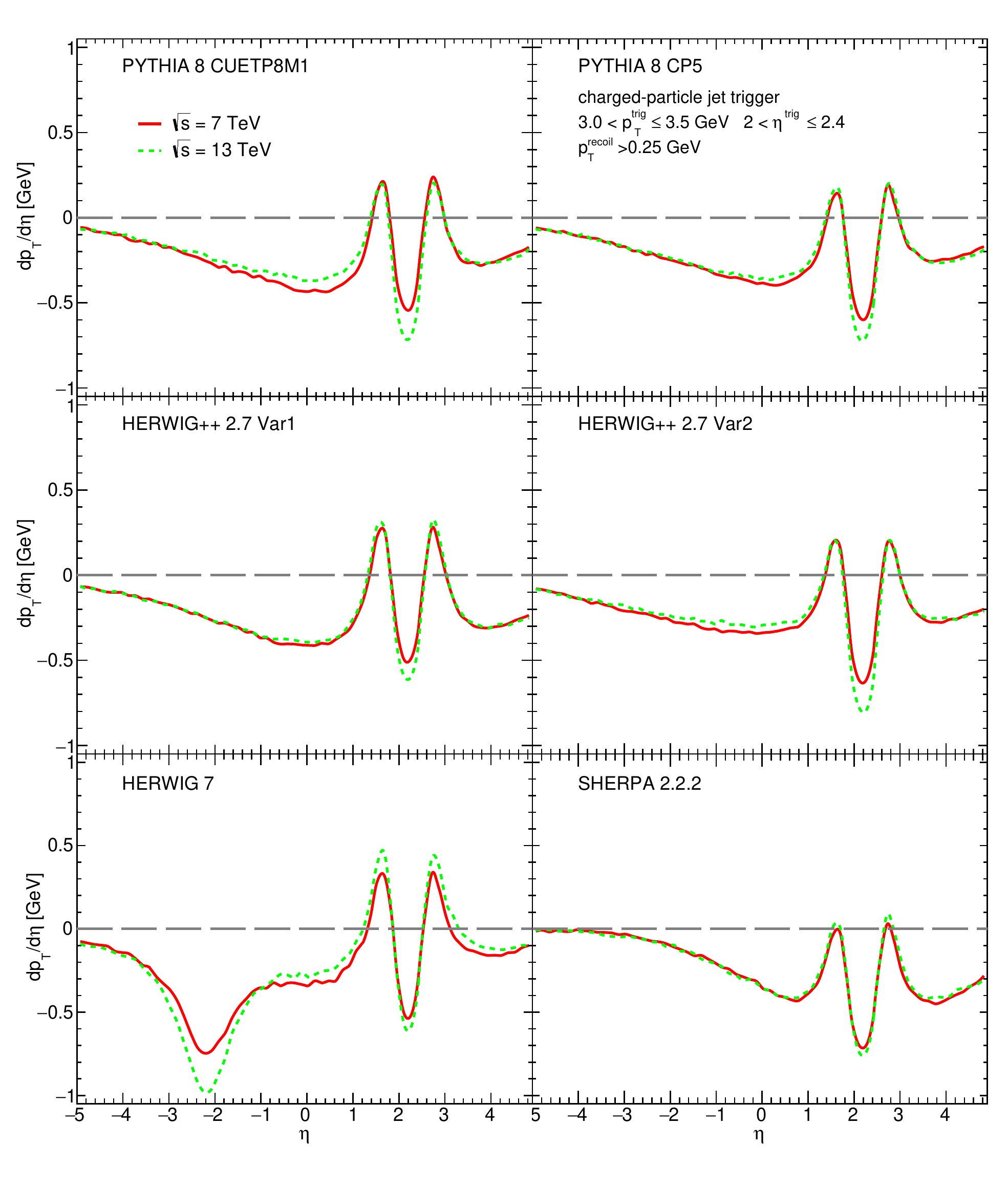}%
\caption{Comparison  of rapidity correlation of recoiled system with respect to charged-particle jet
with $3  \leq p_{\rm T}< 3.5$~GeV  at $\sqrt{s}=~7$~TeV and $\sqrt{s}=13$~TeV for different models. }%
\label{RapidityCorrJet7vs13TeV}%
\end{center}
\end{figure}

Let us turn to the discussion of the dependence of the rapidity correlations on the collision energy. 
Fig.~\ref{RapidityCorrJet7vs13TeV} shows a comparison of the observable at  $\sqrt{s}=7$~TeV and $\sqrt{s}=13$~TeV.
All presented models, except \sherpa, show significant increase yield of particles associated with the trigger jet with the increase 
of the collision energy. 
However, for all the models the $p_{\rm T}^{\rm rec}$ distributions 
converge in the distant $\eta$ regions.
In both \herwig and \pythia the only explicitly energy-dependent parameter is $p_{T0}(s)$, see Eq.~(\ref{eq:ptevol}).
However, indirectly, the energy evolution is also encoded for example in the $x$ dependence of the PDFs.
In the top two panels of Fig~\ref{RapidityCorrJet7vs13TeV}, we show results for the two \pythia tunes. From the plots 
it is clear that the evolution with  $\sqrt{s}$ is stronger for the CP5 tune, which has a larger value 
of $p_{T0}(13~\textrm{TeV})=2.8$ GeV, compare to tune CUETP8M1 which has $p_{T0}(13~\textrm{TeV})=1.44$ GeV.  
The $p_{T0}(s)$ dependence of \pythia prediction seems to be consistent with trends which we studied in more details in
 Appendix~\ref{Sec:Correlations}.
It is also interesting to notice that in CP5 tune the power $\lambda$ governing the $p_{T0}(s)$ evolution is equal 
to $0.03344$ meaning the parameter is almost energy independent. Therefore, the energy dependence of CP5 tune is mainly
governed by the PDF. On the other had in \herwig, we observe the opposite trend, tune Var2 which has smaller 
$p_{T0}(13~\textrm{TeV})$ then tune Var1 (see Appendix B), shows stronger energy dependence, 
see middle panels of Fig~\ref{RapidityCorrJet7vs13TeV}. Finally, \sherpa model predicts, to good approximation, 
no energy dependence for the observable. 
%

\begin{figure}[hbtp]
\begin{center}
\includegraphics[ width=1.0\textwidth]{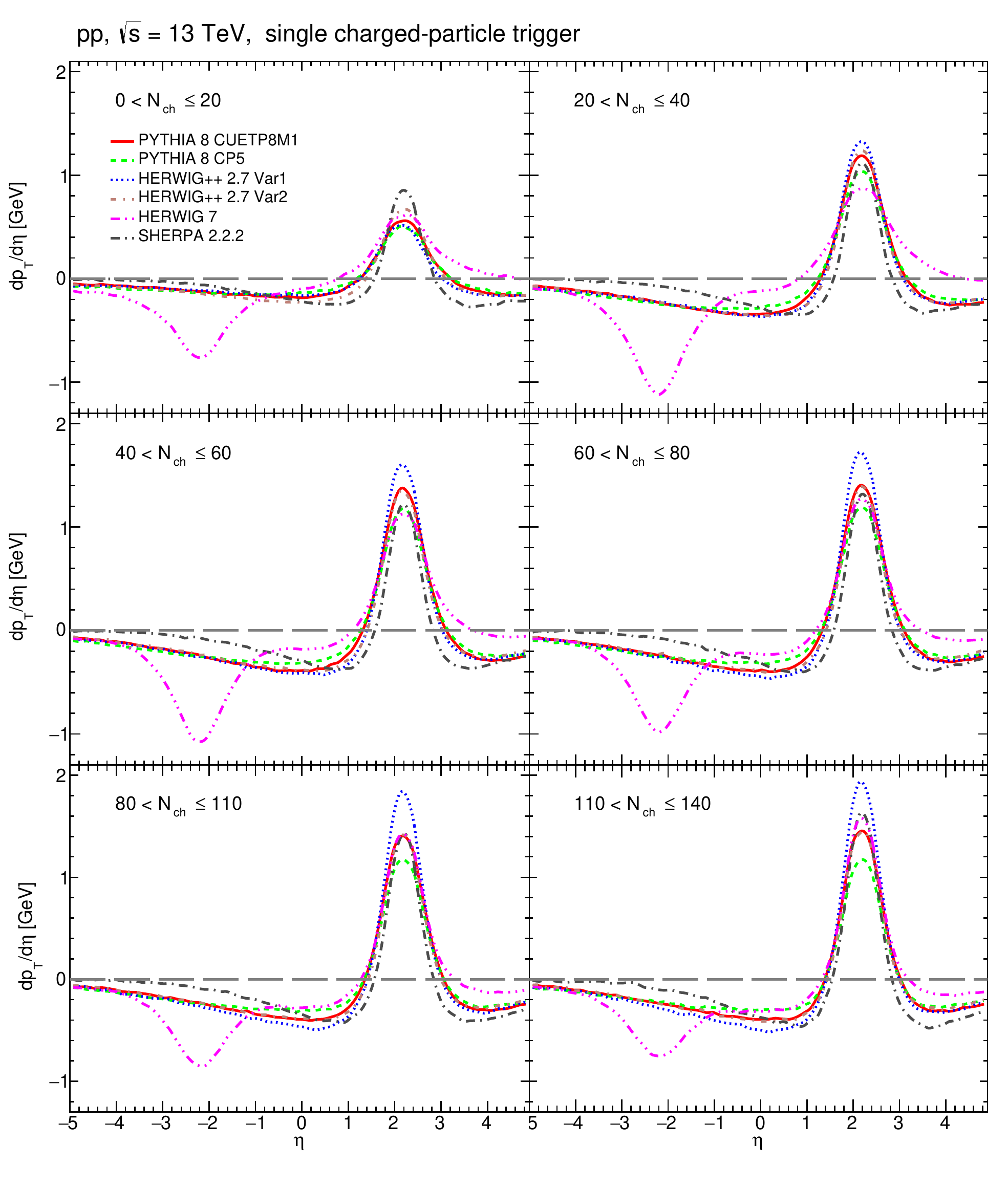}%
\caption{ Comparison  of rapidity correlation of recoiled system with respect to single charged particle 
with $1.5  \leq p_{\rm T}< 2$~GeV  at $\sqrt{s}=~13$~TeV in different $N_{\rm ch}$ domains for various MC models.}%
\label{RapidityCorrMult}%
\end{center}
\end{figure}

The study of correlations for different intervals of $N_{\rm ch}$ reflects the transverse structure of colliding protons. 
The $N_{\rm ch}$ is defined here as a number of stable charged particles with $p_{\rm T}>$~250~MeV 
and $|\eta|<2.4$. We choose a single charged particle as the trigger for the present study.
This is motivated by an increase of the UE contribution into a jet cone with increasing $N_{\rm ch}$. For instance,
the average $p_{\rm T}$ density of charged particles is  roughly 3~GeV per square unit 
at $N_{\rm ch}=100$. That can contribute as much as half of the trigger jet momentum.
In Fig.~\ref{RapidityCorrMult} we show the comparison of the $p_{\rm T}^{\rm rec}$ distribution
in different $N_{\rm ch}$ domains at $\sqrt{s}=$~13~TeV for various MC models. 
\pythia and \herwig models exhibit fast increase of the amplitude of the rapidity correlation, up to  $N_{\rm ch} \approx 60$, and then it saturates. 
Such behaviour mimics the transverse $p_{\rm T}$ density in the UE analyses~\cite{UE_CMS_7000,UE_ALICE_7000, UE_ATLAS_7000}.
At the same time \sherpa model shows continuous  increase of the peak along the trigger direction. 
The difference between the two \herwig{++ 2.7} tunes could be explained by the difference
in the transverse proton structure, which is conventionally characterized by $\sigma_{\rm eff}$.
The peak along the trigger particle is expected to be higher for lower $\sigma_{\rm eff}$ and indeed
this is confirmed by the \herwig{++} results in Fig.~\ref{RapidityCorrMult}, where the 
tune Var1 has smaller $\sigma_{\rm eff}$ ($14.8$~mb) compared to the tune Var2 ($20.6$~mb). 
However, interestingly, there is an opposite trend for the 
\pythia tunes CUETP8M1 and CP5. The $\sigma_{\rm eff}$ for CUETP8M1 is 27.9~mb~\cite{Khachatryan:2015pea}, 
while for CP5 $\sigma_{\rm eff}$ is 25.3~mb~\cite{CMS:2018zub}, as obtained 
in the inclusive 4-jet production.
Therefore, the observed difference is probably related to the change of PDF in the both \pythia tunes.

\begin{figure}[hbtp]
\begin{center}
\includegraphics[ width=0.5\textwidth]{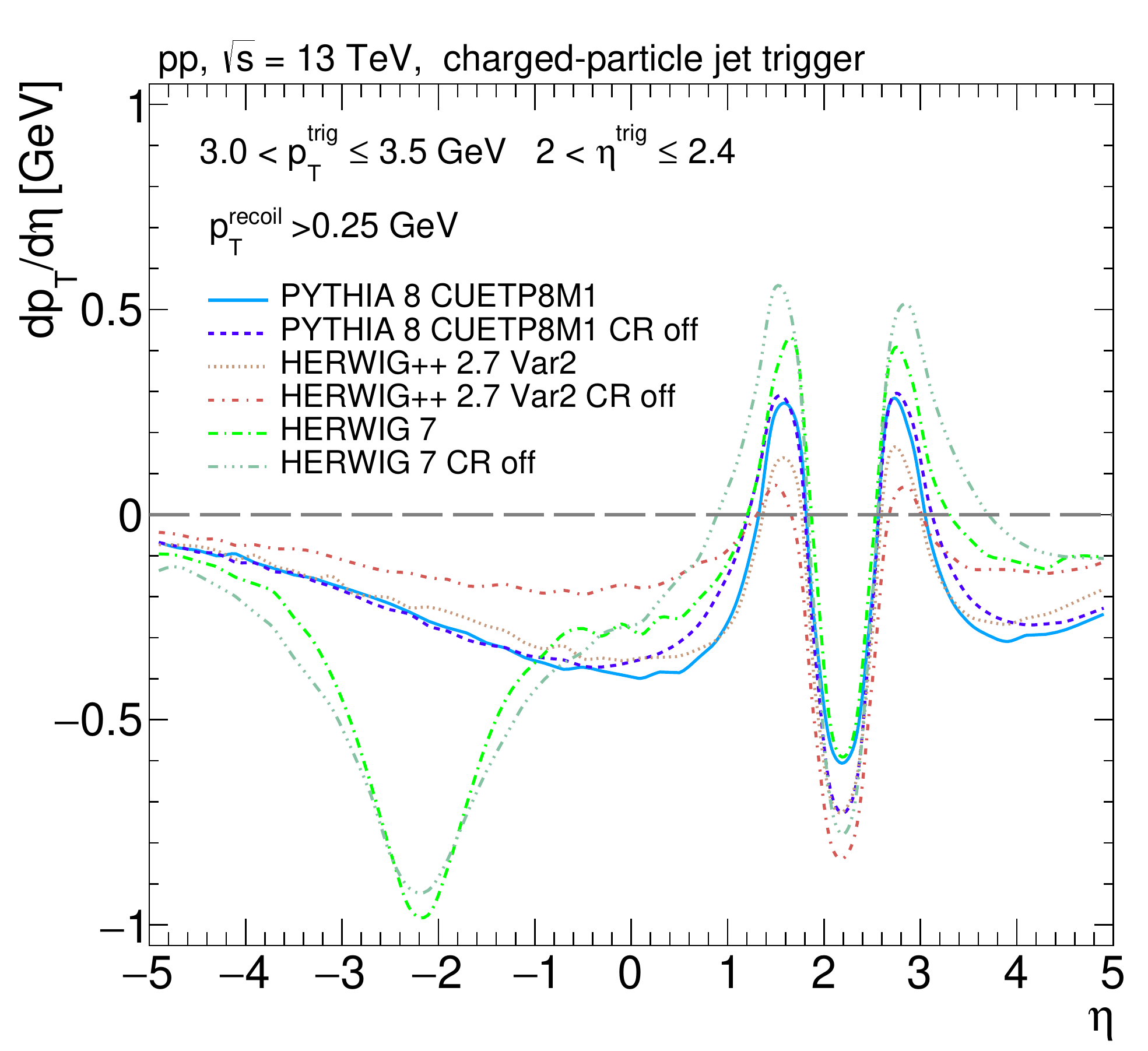}%
\caption{ Effect of color reconnection in \pythia 8 and \herwig{++ 2.7} models. }%
\label{CR_effect}%
\end{center}
\end{figure}

The color reconnection mechanism is one of the least understood elements of MPI models. 
Therefore, it is natural to test whether the proposed observable
is sensitive to the CR. Fig.~\ref{CR_effect} shows  results of  switching the CR on and off 
in \pythia~8 and \herwig models. The former shows most significant differences
close the trigger region, while they almost converge at $\eta= -1$. 
For the \herwig{++ 2.7} the effect is  qualitatively similar, however the versions with CR 
and CR-off do not converge within the studied $\eta$-range. Such behaviour can be qualitatively 
explained  by the fact that the trigger jet ``absorbs''  softer jets during the CR procedure.  

\begin{figure}[]
\begin{minipage}[h]{0.49\linewidth}
\center{\includegraphics[width=1\linewidth]{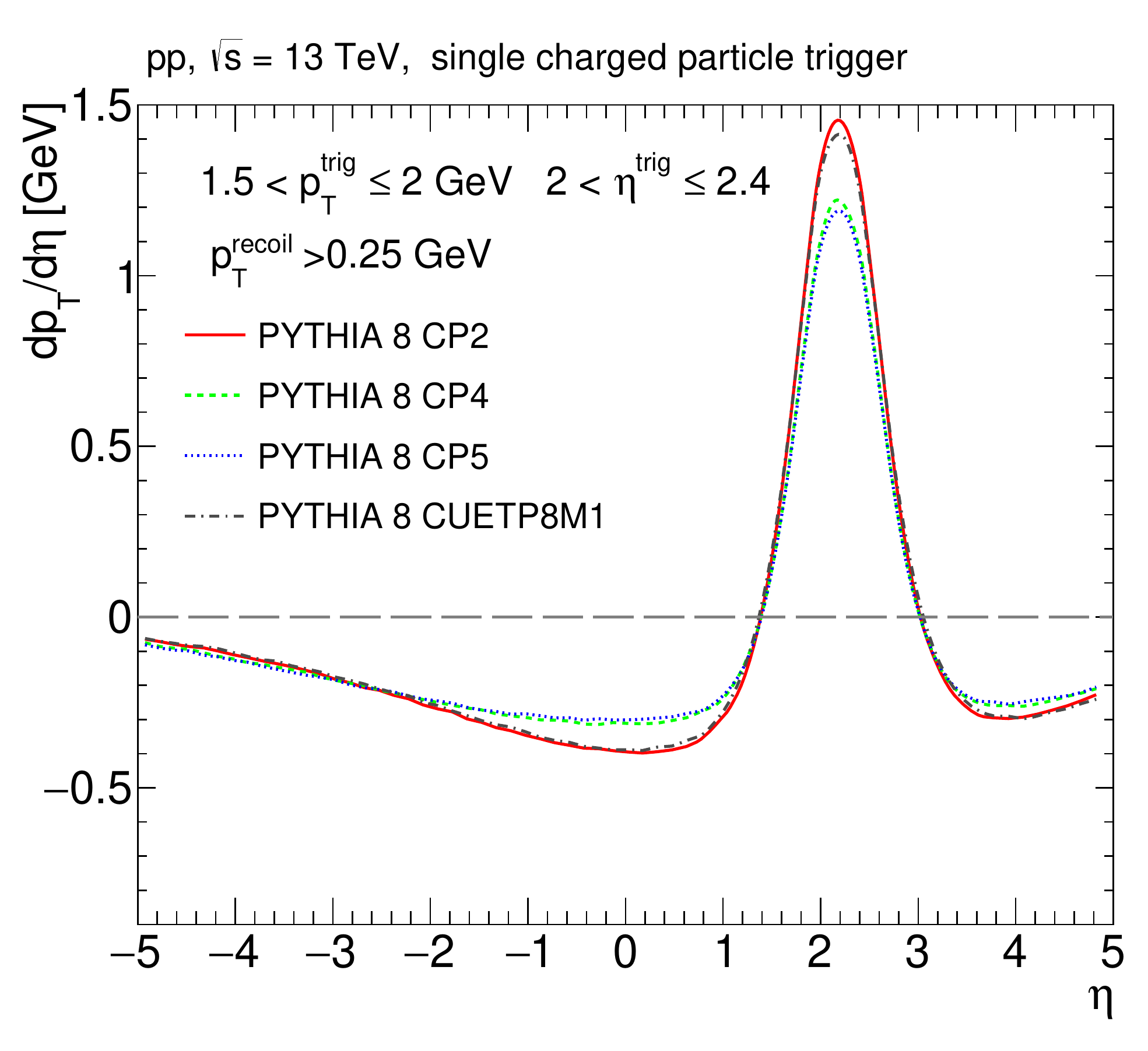} \\ (a)}
\end{minipage}
\begin{minipage}[h]{0.49\linewidth}
\center{\includegraphics[width=1\linewidth]{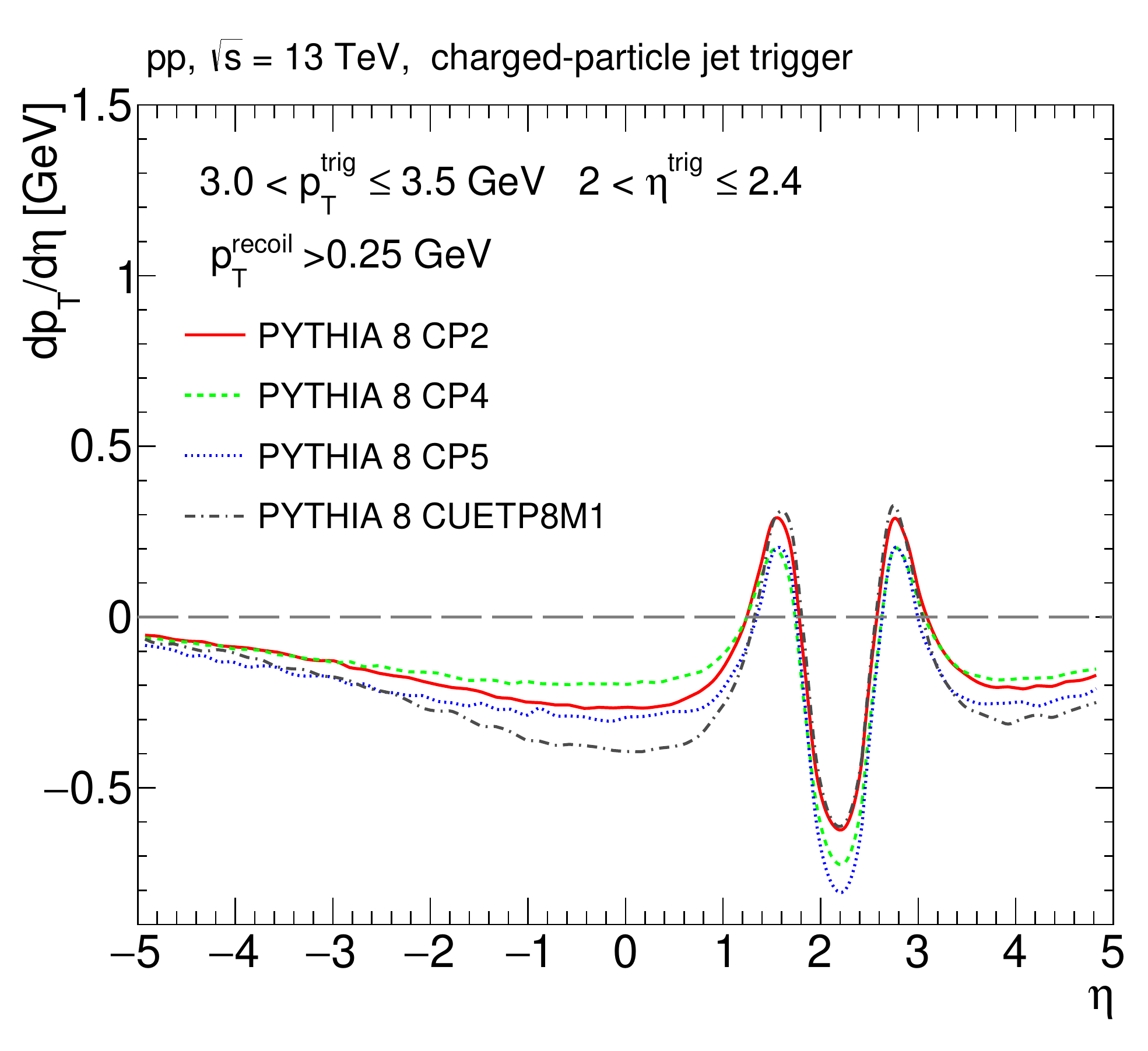} \\ (b)}
\end{minipage}

\caption{Rapidity correlation of recoiled system with respect to single-charged particle 
with 1.5 $ \leq p_{\rm T}<$ 2 GeV (a) and with respect to charged-particle jet
with 3 $ \leq p_{\rm T}<$ 3.5 GeV (b) at $\sqrt{s}=$~13~TeV.}%
\label{RapidityCorr_PythiaOnly}
\end{figure}

It is also instructive to compare the expectations of several models within a single event generator for the proposed observable. 
For this purpose we choose \pythia 8 tunes  discussed  above and others presented in the ~\cite{CMS:2018zub}, 
namely \pythia 8 CUETP8M1, CP2, CP4, CP5. The key features of these tunes are discussed in Section~\ref{sec:PythiaModel}.
Here, let us only remind that the first and second tunes use LO PDF sets, while latter are based on NNLO PDF sets.   
The main difference between CUETP8M1 and CP2 is that for the latter the rapidity odering for the ISR is switched off.
Similar difference is for CP4 and CP5, respectively. In order to study a sensitivity of the proposed observable 
to various physics mechanisms to the maximum extent, we use both discussed approaches to a trigger object.
Fig.~\ref{RapidityCorr_PythiaOnly} (a) clearly shows that the usage of a single charged particle as a trigger makes
the proposed observable sensitive to the choice of PDF sets mainly. The MC tunes in this case can be grouped according to PDF sets,
i.e. LO  and NNLO ones. The difference within a single group is almost within the line width in Fig.~\ref{RapidityCorr_PythiaOnly}.
The difference between the models look  very different if a charged-particle jet is used as
a trigger. However, Fig.~\ref{RapidityCorr_PythiaOnly} (b) shows that MC models can be still grouped by 
a choice of PDF sets.  The depth of the minimum is very close within the single PDF set. 
The particles that have $\eta<1$  give a higher recoil in models with rapidity odering during development
of initial state shower.

%% file: Summary.tex
We have introduced a new observable which probes interplay between the soft and hard physics at moderate $p_{\rm T}$
via probing  long and short range rapidity correlations of transverse momenta of charged particles/minijets.
The basic idea is to study how the transverse momenta of hadrons produced in association with a trigger object are balanced 
as a function of rapidity (the precise definition is given in Section~\ref{Sec:Observables}).
It is shown that the observable is sensitive to basic mechanisms and components used in the present
MC models, such as a suppression of low-$p_{\rm T}$ jet production, parton distribution functions,
a transverse geometry of proton, a color reconnection mechanism, and their evolution with collision energy.  
We demonstrated that predictions of different MC models which describe well many characteristics of 
the hadron production at LHC differ significantly for suggested observable.
The most prominent discrepancy between models appears when the correlation is studied as a function of charged-particle multiplicity.
It is important to stress that changing the parameters within a single model results in the expected changes for the measured
distribution. Therefore, the proposed measurements can help to disentangle various mechanisms relevant for minijet production. 
It is worth also mentioning that our tests have revealed quite peculiar features of \herwig 7 and \sherpa 2.2.2 models.  
 
We performed our tests taking into account performance of general purpose detectors at LHC such as ATLAS and CMS.
Hence, one may hope that prompt experimental studies of the quantities we calculated will be possible.
The data necessary for proposed study are available from low pileup LHC runs.
Standard amounts of minimum bias data, that are usually few ten million events, are enough for the measurement,
however a special trigger is desired.  

Obviously, the discussed correlations are sensitive to the various collective effects.
Hence, it would be also interesting to study such correlations also in $pA$ and $AA$ scatterings. 
The proposed  measurements can be also extended by using as trigger particles two hadrons with  
azimuthal angle difference $\phi_1 - \phi_2\sim \pi/2$. One would measure $\langle k(\phi_1,\Delta y_1) \rangle $
and $\langle  k(\phi_2,\Delta y_2) \rangle $ and compare the results with the measurements of the same quantities with 
one trigger particle which we studied in this paper. 
We  expect that such an observable would have an enhanced  sensitivity to the contribution of the multiparton interactions and collective effects.

\section*{Acknowledgments}
We would like to thank Frank Krauss for reading the \sherpa{} model Section. 
Mark Strikman would like to thank CERN for hospitality, where this work has started.
The research of Mark Strikman and partially of Piotr Kotko was supported by the U.S. Department of Energy,
Office of Science, Office of Nuclear Physics, under Award No. DE-FG0 2-93ER40771. The research of Piotr Kotko was also supported by
the U.S. Department of Energy grant No. DE-SC-0002145. Andrzej Siodmok acknowledges support from
the National Science Centre, Poland Grant No. 2016/23/D/ST2/02605 and the European Union’s Horizon 2020 research
and innovation programme as part of the Marie Sklodowska-Curie Innovative Training Network MCnetITN3
(grant agreement no. 722104).

%% file: Appendix.tex
\input{PythiaCorr.tex}

\newpage

\section{Herwig++ Parameters}
\FloatBarrier

\begin{table*}[!hbt]
\begin{center}
  \begin{tabular}{lccc} \toprule & \textit{Only UE data in fit (Var2)} & \textit{ UE data and \seff in fit (Var1)} \\
    \midrule
    $\mu^2\enspace [GeV^2]$       & 1.65  &  2.30    \\
    $\pdisrupt$                    & 0.22  &  0.80    \\
    $\preco$                          & 0.60  &  0.49    \\
    \midrule
    $\ptminnought\enspace [GeV]$      & 2.80  &  3.91    \\
    $b$                                & 0.29  &  0.33    \\
    \bottomrule
  \end{tabular}
  \caption{Parameters of the underlying event tunes. The
  last two parameters describe the running of $\ptmin$ according to
  Eq.~(\ref{eq:ptevol}).}
  \label{tab:fixedparams}
 \end{center}
\end{table*}
\FloatBarrier

\section{SHERPA Parameters}
\label{Shepa-app}
\FloatBarrier

\begin{table*}[!hbt]
\begin{center}
  \begin{tabular}{lcc} parameter & value \\
    \midrule
    $SOFT\_COLLISIONS $       & Shrimps    \\
    $Shrimps\_Mode$           & Inelastic  \\
    $\Delta Y$                &  1.50    \\
    $\Lambda^2$               &  1.376   \\
    $\beta_0^2 $              &  18.76   \\
    $\kappa$                  &  0.6     \\
    $\xi$                     &  0.2     \\
    $\lambda$                 &  0.2151  \\
    $\Delta$                  &  0.3052  \\
    $Q_0^2$                   &  2.25    \\
    $\chi_{\rm S}$                  &  1.0     \\
    $Shower\_Min\_K_{\rm T}^{2}$  &  4.0     \\
    $Diff\_Factor$            &  4.0     \\
    $K_{\rm T}^{2}\_Factor$       &  4.0     \\
    $RescProb$                &  2.0     \\
    $RescProb1$               &  0.5     \\
    $Q_{\rm RC}^{2}$              &  0.9     \\
    $ReconnProb$              &  -25     \\
    $Resc\_K_{\rm T,Min}$        &  off     \\
    $Misha$                   &  0       \\
    \bottomrule
  \end{tabular}
  \caption{Parameters of the \sherpa 2.2.2 model for the production of minimum bias events.}
  \label{tab:sherpa_minbias}
 \end{center}
\end{table*}
\FloatBarrier

%% file: PythiaCorr.tex
\section{Minijet correlations in \pythia}
\label{Sec:Correlations}
 
\FloatBarrier 

It is instructive to study how the $\left\langle p_{\rm T}^{\rm rec}\right\rangle$ 
distribution defined in Section~\ref{Sec:Observables} depends on the crucial parameters which modify the way minijets and MPI are
generated. 
 Recall, that in the most simple MPI model (with the hard collisions  completely
uncorrelated) the contribution to $\left\langle p_{\rm T}^{\rm rec}\right\rangle$ from independent sub-systems cancels out. 
Hence, there survives only the  contribution the system to which the trigger belongs.
Since in \pythia correlations are present, the distribution will be sensitive to the MPI mechanism.
For the purpose of this study, we choose one of the standard \pythia tunes (Monash~\cite{Skands:2014pea}), 
in which we will play with MPI on/off feature and modify the $p_{\rm T0}$ parameter.

\begin{figure}
\begin{centering}
\includegraphics[width=9cm]{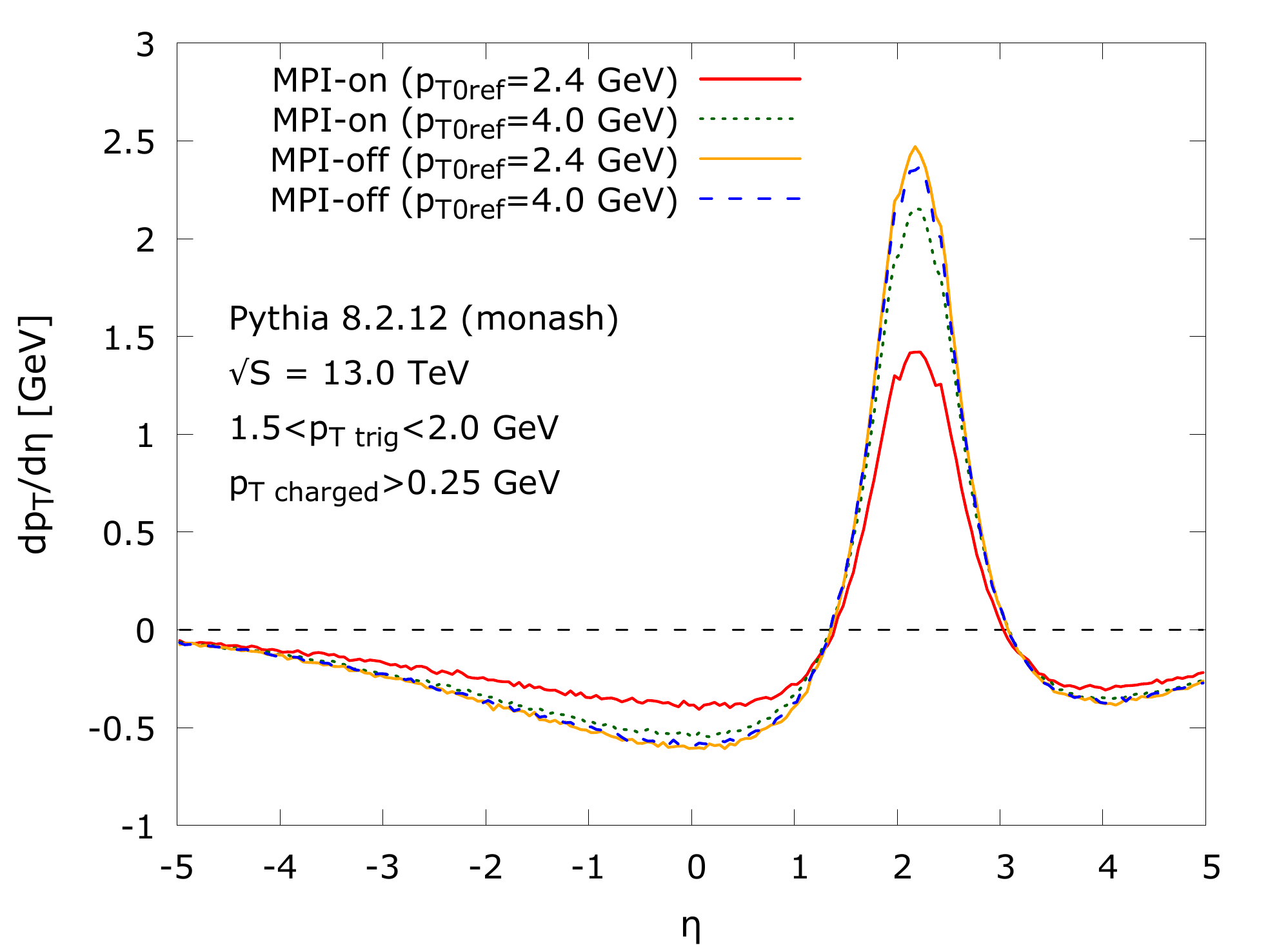}
\par\end{centering}

\caption{$\left\langle p_{T}^{\mathrm{rec}}\right\rangle $ as a function of
rapidity in \pythia with hadronization, full beam remnant treatment,
initial and final state showers. We study effect of MPI on/off in
the simulation. We also change the standard parameter for the $p_{T0}$
($2.4\,\mathrm{GeV}$) cutoff in the MPI model to higher value ($4.0\,\mathrm{GeV}$)
to observe how this affects the distribution. \label{fig:pTmean_MPI}}
\end{figure}

The result with MPI feature on and off is presented in Fig.~\ref{fig:pTmean_MPI}.
We also studied the effect of changing the  $p_{\rm T0}$ parameter. 
The simulation was performed with full hadronization. In order to make it possible to connect the present simulations 
with our main results in Section~\ref{Sec:Results},  we used only charged
particles and required all particles to have $p_{\rm T}>0.25\,\mathrm{GeV}$. 
Removal of very soft charged particles does  not change the conclusions. 
First, we observe that the change of $p_{\rm T0}$ for the no-MPI scenario
has a very little effect. This was already observed in Section~\ref{sec:BasicMinijetModel}
for the basic perturbative minijet model. Second, for the standard $p_{\rm T0}=2.4\,\mathrm{GeV}$
the effect of turning the MPI feature on is dramatic. The distribution
is scaled down by a factor of about $0.6$. If the $p_{\rm T0}$ cutoff
is raised to $4\,\mathrm{GeV}$, the scaling factor is only about $0.9$.

\begin{figure}
\begin{centering}
\includegraphics[width=9cm]{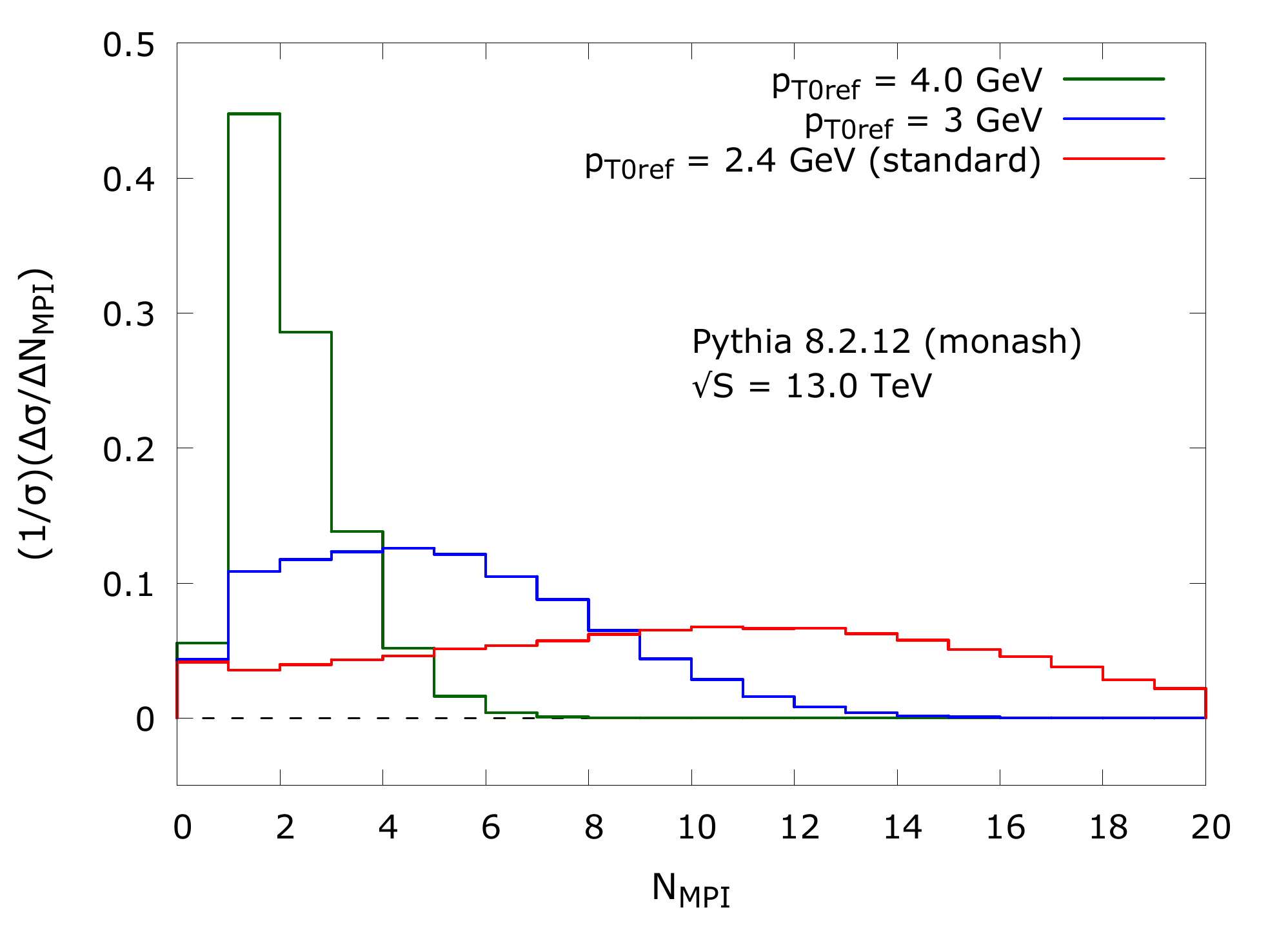}
\par\end{centering}

\caption{The distribution of number of parton interactions $N_{\mathrm{MPI}}$
for different settings of $p_{T0}$ cutoff. The events with $N_{\mathrm{MPI}}=0$
are diffractive events.\label{fig:MPIdistrib}}
\end{figure}

The above results suggest that: (i) the distribution $\left\langle p_{T}^{\mathrm{rec}}\right\rangle $
is very sensitive to the MPI (at least in the \pythia model), (ii)
the $p_{T0}$ cutoff is tightly connected to the number of MPI generated
(as one should expect). The second point can be directly illustrated
by an explicit calculation. In Fig.~\ref{fig:MPIdistrib}
we show how the mean value of the MPI number changes when
we change $p_{T0}$. We see that for the Monash tune with $p_{T0}=2.4\,\mathrm{GeV}$
the average number of MPI is more than 10. For $p_{T0}=4.0\,\mathrm{GeV}$
it narrows down to something between 1 and 2.

\begin{figure}
\begin{centering}
\includegraphics[width=9cm]{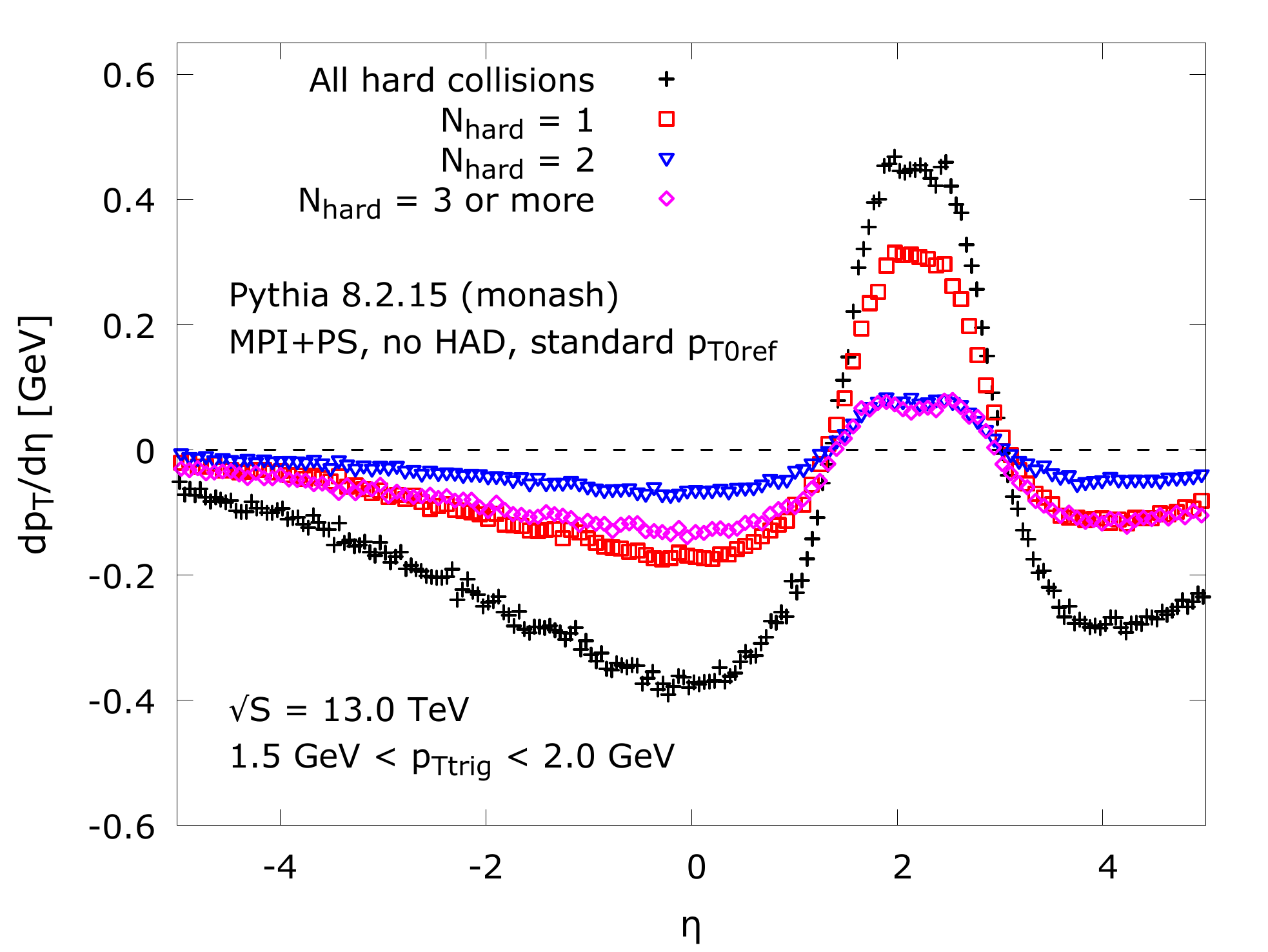}
\par\end{centering}

\caption{$\left\langle p_{\rm T}^{\mathrm{rec}}\right\rangle $ decomposed into
contributions from subsequent (in $p_T$) hard collisions with $N_{\mathrm{hard}}$ =1,2 and 3 or more. The
grouping into hard subsystems was done without hadronization. \label{fig:pTrec_Nhard}}
\end{figure}

\begin{figure}
\begin{centering}
\includegraphics[width=9cm]{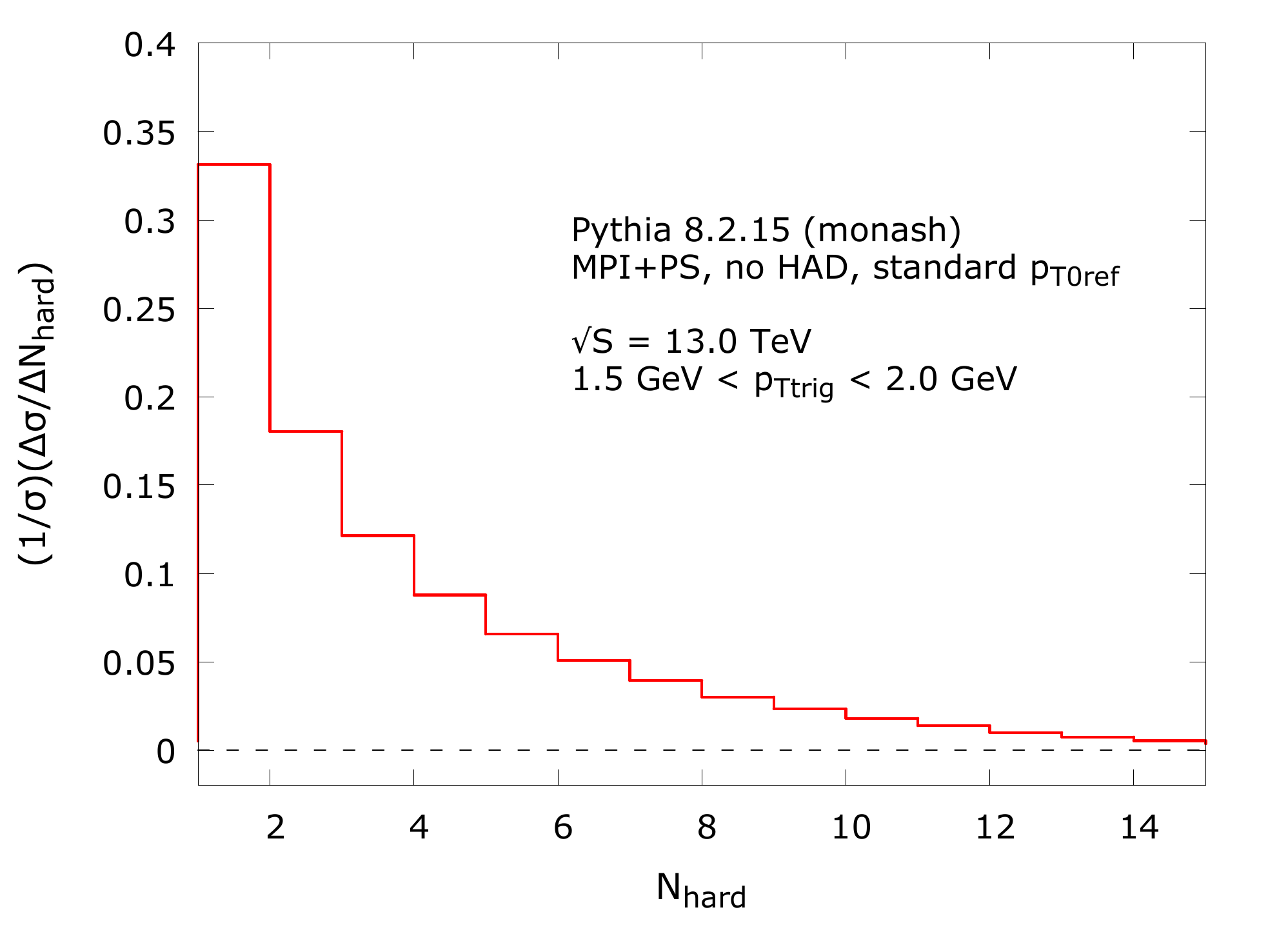}
\par\end{centering}

\caption{Distribution of a trigger particle among the hard subsystems produced by the MPI mechanism. The
subsystems are ordered according to the $p_{T}$ of the hard process.\label{fig:trigger_Nhard}}
\end{figure}

The way the $\left\langle p_{T}^{\mathrm{rec}}\right\rangle $
is sensitive to MPI in \pythia can be understood with the help of the
following calculation. We switch off the hadronization and use 
the algorithm that groups the final states with respect to the parent
hard process. Then we calculate what is the contribution of subsequent
hard collisions to $\left\langle p_{T}^{\mathrm{rec}}\right\rangle$. 
The results are shown in Fig.~\ref{fig:pTrec_Nhard}. We see, that
due to the $p_{T}$ ordering of MPI, the $\left\langle p_{T}^{\mathrm{rec}}\right\rangle $
for the subsequent hard collisions is scaled down more and more. Note,
that if the trigger belonged to the system with $N_{\mathrm{hard}}=i$,
all contributions to $\left\langle p_{T}^{\mathrm{rec}}\right\rangle $
from $N_{\mathrm{hard}}<i$ is zero. Thus, how much the distribution is scaled down, depends on how often the trigger falls into to various hard sub-systems. This is answered explicitly
by the calculation presented in Fig.~\ref{fig:trigger_Nhard}. 
We see that the trigger often originates from the non-hardest parton interaction ($N_{\mathrm{hard}}>$~1)
that scales down the $\left\langle p_{\rm T}^{\mathrm{rec}}\right\rangle $  distribution.
This is a genuine effect of MPI correlations in \pythia caused by ordering of the binary parton collisions. 
The other correlations have weaker effect.

\begin{figure}
\begin{centering}
\includegraphics[width=7cm]{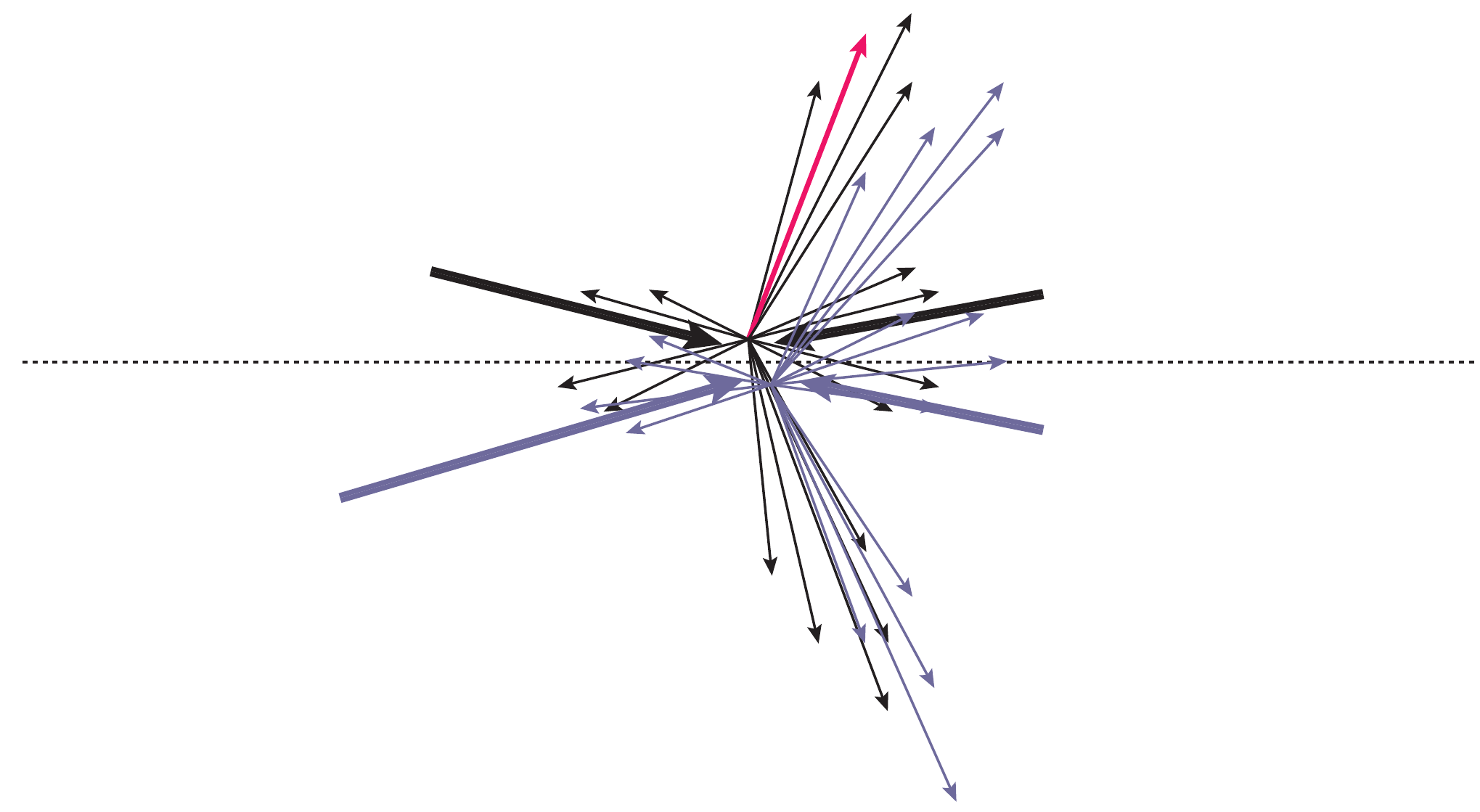}
\par\end{centering}

\caption{Two hard collisions with transverse correlations schematically represented
by the transverse momentum exchange between incoming partons. The
total transverse momentum does not sum up to zero for each subsystem,
but overall is conserved. \label{fig:MPI_transvmom}}
\end{figure}

\begin{figure}
\begin{centering}
\includegraphics[width=9cm]{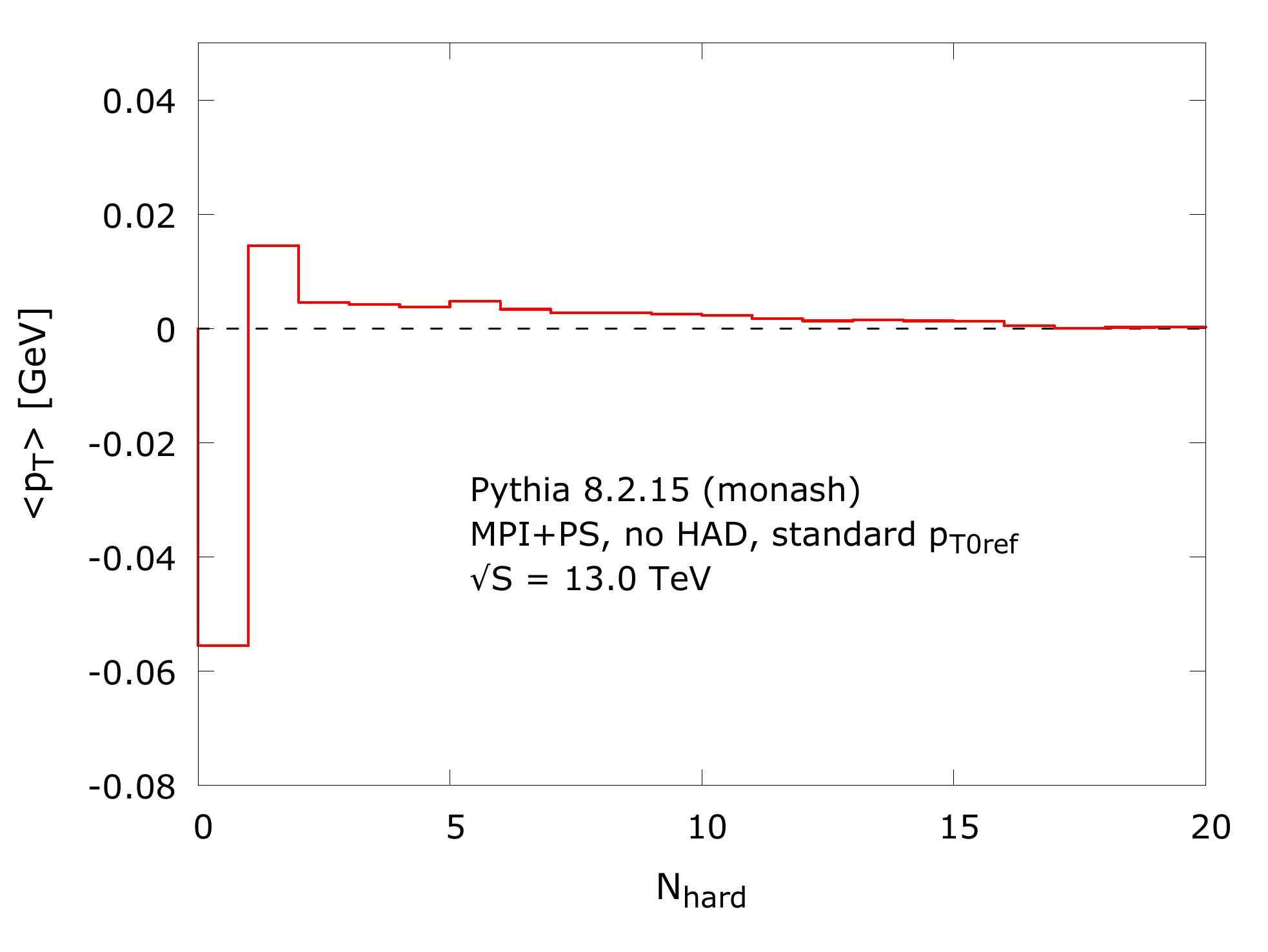}
\par\end{centering}

\caption{Distribution of the mean $p_{T}$ as a function of the number of parton interaction in \pythia.
Although the total transverse momentum is conserved,
the subsequent hard collisions posses a slights transverse imbalance
due to the primordial $k_{T}$ mechanism.\label{fig:MPI_momcons}}
\end{figure}

In \pythia, in general, the transverse momentum is
not conserved in the individual hard parton collisions (but of course
it is conserved for the whole event). The idea is schematically illustrated
in Fig.~\ref{fig:MPI_transvmom}. 
In order to see this explicitly we use again the algorithm to
group the final state particles \textit{before hadronization} into
groups belonging to different hard process and the beam remnants
as a separate class. We switch off the hadronization to slightly simplify the procedure, as tracing the final state hadrons back to the
hard process is not possible in a unique way. The hard collisions are
enumerated $N_{\mathrm{hard}}=1,2,\dots$ from hardest to softest,
with $N_{\mathrm{hard}}=0$ reserved for the beam remnants. 
In Fig.~\ref{fig:MPI_momcons} we show the $\left\langle p_{T}\right\rangle $
(defined as before but now we do include the trigger) as a function
of $N_{\mathrm{hard}}$. We see that indeed there are transverse
momentum correlations between MPI. We check that they are generated
by the primordial $k_{T}$ mechanism, that is if the mechanism is
switched off, the result for $\left\langle p_{T}\right\rangle \left(N_{\mathrm{hard}}\right)$
is approximately $0$ everywhere.
Note, that our observable $\left\langle p_{T}\right\rangle $
is not sensitive to the direct transverse momentum correlations discussed above.

\FloatBarrier